\documentclass{vldb}

\makeatletter
\newif\if@restonecol
\makeatother

\usepackage{url}
\usepackage{times}
\usepackage[vlined,algoruled]{algorithm2e}
\usepackage{verbatim}   
\usepackage{color}      
\usepackage{bm}                     	
\usepackage{enumitem}

\usepackage{caption}
\usepackage{subcaption}
\usepackage{balance}
\usepackage{multirow}

\linesnumbered
\SetKwInOut{Input}{Input}
\SetKwInOut{Output}{Output}
\SetKw{Break}{break}
\SetKw{Continue}{continue}
\SetCommentSty{textsl}

\newcommand{\eat}[1]{}
\newcommand{\tightlist}{\itemsep=-2pt}
\newcommand{\rbox}{\hfill $\Box$}
\newcommand{\ie}{{\em i.e.}}
\newcommand{\eg}{{\em e.g.}}

\newcommand{\SqV}{SqV\xspace}
\newcommand{\SqC}{SqC\xspace}
\newcommand{\SqA}{SqA\xspace}
\newcommand{\SplitAndMerge}{{\sc MultiLayerSM}\xspace}
\newcommand{\SplitAndMergePlus}{{\sc MultiLayerSM+}\xspace}

\newcommand{\dom}{\mbox{dom}}

\newcommand{\argmax}{\operatornamewithlimits{argmax}}

\newcommand{\defeq}{\triangleq}
\newcommand{\ind}[1]{\mathbb{I}(#1)}

\newcommand{\accu}{{\sc Accu}\xspace}
\newcommand{\popaccu}{{\sc PopAccu}\xspace}
\newcommand{\Nam}{N.Amer.\xspace}
\newcommand{\pre}{\mbox{Pre}}
\newcommand{\abs}{\mbox{Abs}}

\newcommand{\smallex}{}

\begin{document}

\numberofauthors{1}
\author{
\alignauthor
Xin Luna Dong, Evgeniy Gabrilovich, Kevin Murphy, Van Dang \\
Wilko Horn, Camillo Lugaresi, Shaohua Sun, Wei Zhang\\
 \affaddr{Google Inc.}
 \email{\{lunadong$\mid$gabr$\mid$kpmurphy$\mid$vandang$\mid$wilko$\mid$camillol$\mid$sunsh$\mid$weizh\}@google.com}
}

\title{Knowledge-Based Trust: Estimating \\the Trustworthiness of Web Sources}
\maketitle

\begin{abstract}
The quality of web sources has been traditionally evaluated using {\em exogenous}
signals such as the hyperlink structure of the graph.
We propose a new approach
that relies on {\em endogenous} signals, namely, the correctness of
factual information provided by the source.
A source that has few false facts is considered to be trustworthy.

The facts are automatically extracted from each source
by information extraction methods commonly
used to construct knowledge bases.
We propose a  way to distinguish errors made in the
extraction process from factual errors in the web source per se, by using joint inference
in a novel multi-layer probabilistic model.

We call the trustworthiness score we computed {\em Knowledge-Based Trust (KBT)}.
On synthetic data, we show that our method can reliably compute the
true trustworthiness levels of the sources.
We then apply it to  a database of 2.8B facts extracted from the
web, and thereby estimate the trustworthiness of 119M webpages.
Manual evaluation of a subset of the results confirms the effectiveness of
the method.
\end{abstract}

\newtheorem{definition}{Definition}[section]
\newtheorem{proposition}[definition]{Proposition}
\newtheorem{lemma}[definition]{Lemma}
\newtheorem{remark}[definition]{Remark}
\newtheorem{corollary}[definition]{Corollary}
\newtheorem{claim}[definition]{Claim}
\newtheorem{theorem}[definition]{Theorem}
\newtheorem{example}[definition]{Example}

\newtheorem{review}{Comment}[subsection]
\newcommand{\answer}[1]{{\bf Answer:} #1}

\section{Introduction}
\label{sec:intro}
\vspace{-0.1in}
\begin{quote}
{\em ``Learning to trust is one of life's most difficult tasks.''}
-- Isaac Watts.
\end{quote}

Quality assessment for web sources\footnote{
We use the term ``web source'' to denote a specific webpage,
such as \url{wiki.com/page1},
or a whole website, such as \url{wiki.com}.
We discuss this distinction in more detail in Section~\ref{sec:hierarchy}.
} %
is of tremendous
importance in web search. 
It has been traditionally evaluated using exogenous 
signals such as hyperlinks and browsing history. However, such signals
mostly capture how popular a webpage is.
For example, the  gossip websites listed in~\cite{gossip}
mostly have high PageRank scores~\cite{pagerank}, but would not generally be considered
reliable. Conversely, some less popular websites
nevertheless have very accurate information.

In this paper, we address the fundamental question of estimating how
trustworthy a given web source is.
Informally, we define the trustworthiness or {\em accuracy} of a web source as the
probability that it contains the correct value for a fact (such as
Barack Obama's nationality), assuming that it mentions any value 
for that fact. (Thus we do not penalize sources that have few facts,
so long as they are correct.)
 
We propose using {\em Knowledge-Based Trust (KBT)} to estimate source trustworthiness as follows.
We extract a plurality of facts from many pages using information
extraction techniques. 
We then jointly estimate the correctness of these facts
and the accuracy of the sources 
using inference in a probabilistic model.
Inference is an iterative process, since we believe a source is
accurate if its facts are correct, and we believe the facts are
correct if they are extracted from an accurate source.
We leverage the redundancy of information on the web to break the
symmetry.
Furthermore, we show how to initialize our estimate of the accuracy 
of sources based on authoritative information, in order to ensure that this iterative process
converges to a good solution.

The fact extraction process we use is based on the \emph{Knowledge
Vault} (KV) project \cite{DGH+14}.
KV uses 16 different information
extraction systems to extract (subject, predicate, object)
\emph{knowledge triples} from webpages.
An example of such a triple is
{\em (Barack Obama, nationality, USA)}. A subject 
represents a real-world entity, identified by an ID such as {\em mid}s in {\em Freebase}~\cite{freebase};
a predicate is pre-defined in {\em Freebase},
describing a particular attribute of an entity;
an object can be an entity, a string, a numerical value, or a date.

The facts extracted by automatic methods such as KV may be wrong.
One method for estimating if they are correct or not was described
in \cite{DGE+14}.
However, this earlier work did not
distinguish between factual errors on the page and errors
made by the extraction system. As shown in \cite{DGE+14},
extraction errors are far more prevalent than source errors. Ignoring this
distinction can cause us to incorrectly distrust a website.

Another problem with the approach
used in \cite{DGE+14} is that it estimates the reliability of each
webpage independently. This can cause problems when data are sparse.
For example, for more than one billion webpages,
KV is only able to extract a single triple (other extraction systems
have similar limitations).
This makes it difficult to reliably estimate the trustworthiness of
such sources.
On the other hand, for some pages KV extracts tens of thousands of
triples, which can create computational bottlenecks. 

The KBT method introduced in this paper overcomes some of these previous weaknesses.
In particular, our contributions are threefold.
Our main contribution is a more sophisticated probabilistic model,
which can distinguish between two main sources of error:
incorrect facts on a page, and incorrect extractions made by an
extraction system. This provides a much more accurate estimate of
the source reliability. We propose an efficient, scalable algorithm  for
performing inference and parameter estimation in the proposed probabilistic model
(Section~\ref{sec:fusion}).

Our second contribution is a new  method to adaptively decide the
granularity of sources to work with: if a specific webpage
yields too few triples, we may aggregate it with other webpages from the same website.
Conversely, if a website has too many triples, we may
split it into smaller ones, to avoid computational bottlenecks
(Section~\ref{sec:hierarchy}).

The third contribution of this paper is a detailed, large-scale evaluation of 
the performance of our model. In particular, we applied it to 2.8 billion
triples extracted from the web, and were thus able to reliably predict the trustworthiness
of  119 million webpages and 5.6 million websites (Section~\ref{sec:exp}).

We note that source trustworthiness provides an additional signal for evaluating
the quality of a website. We discuss new research opportunities 
for improving it and using it in conjunction with
existing signals such as PageRank (Section~\ref{sec:disc}).
Also, we note that although we present our methods in the context of
knowledge extraction,
the general approach we propose can be applied to many other tasks
that involve data integration and data cleaning.

\section{Problem Definition and Overview}
\label{sec:definition}
{\small
\begin{table}[t]
\vspace{-.1in}
\centering
\caption{Summary of major notations used in the paper. \label{tbl:notation}}
\vspace{-.1in}
\begin{tabular}{c|l}
    \hline
    \textbf{Notation} & \multicolumn{1}{c}{\textbf{Description}}\\
    \hline
    $w \in \cal W$    & Web source \\
    $e \in \cal E$    & Extractor \\
    $d$        & Data item \\
    $v$        & Value \\
    \hline
    $X_{ewdv}$   & Binary indication of whether $e$ extracts $(d,v)$ from $w$ \\
    $X_{wdv}$    & All extractions from $w$ about $(d,v)$ \\
    $X_{d}$     & All data about data item $d$ \\
    $X$         & All input data \\
    \hline
    $C_{wdv}$    & Binary indication of whether $w$ provides $(d,v)$ \\
    $T_{dv}$    & Binary indication of whether $v$ is a correct value for $d$ \\
    $V_{d}$     & True value for data item $d$ under single-truth assumption \\
    $A_{w}$     & Accuracy of web source $w$ \\
    $P_{e}, R_{e}$  & Precision and recall of extractor $e$ \\
    \hline
\end{tabular}
\vspace{-.1in}
\end{table}
}

In this section, we start with a formal definition of {\em
  Knowledge-based trust} (KBT).
We then briefly review our prior work that solves a closely related problem, 
{\em knowledge fusion} \cite{DGE+14}.
Finally, we give an overview of our approach, and summarize the
difference 
from our prior work.

\begin{figure}[t]
\center
\includegraphics[scale=0.35]{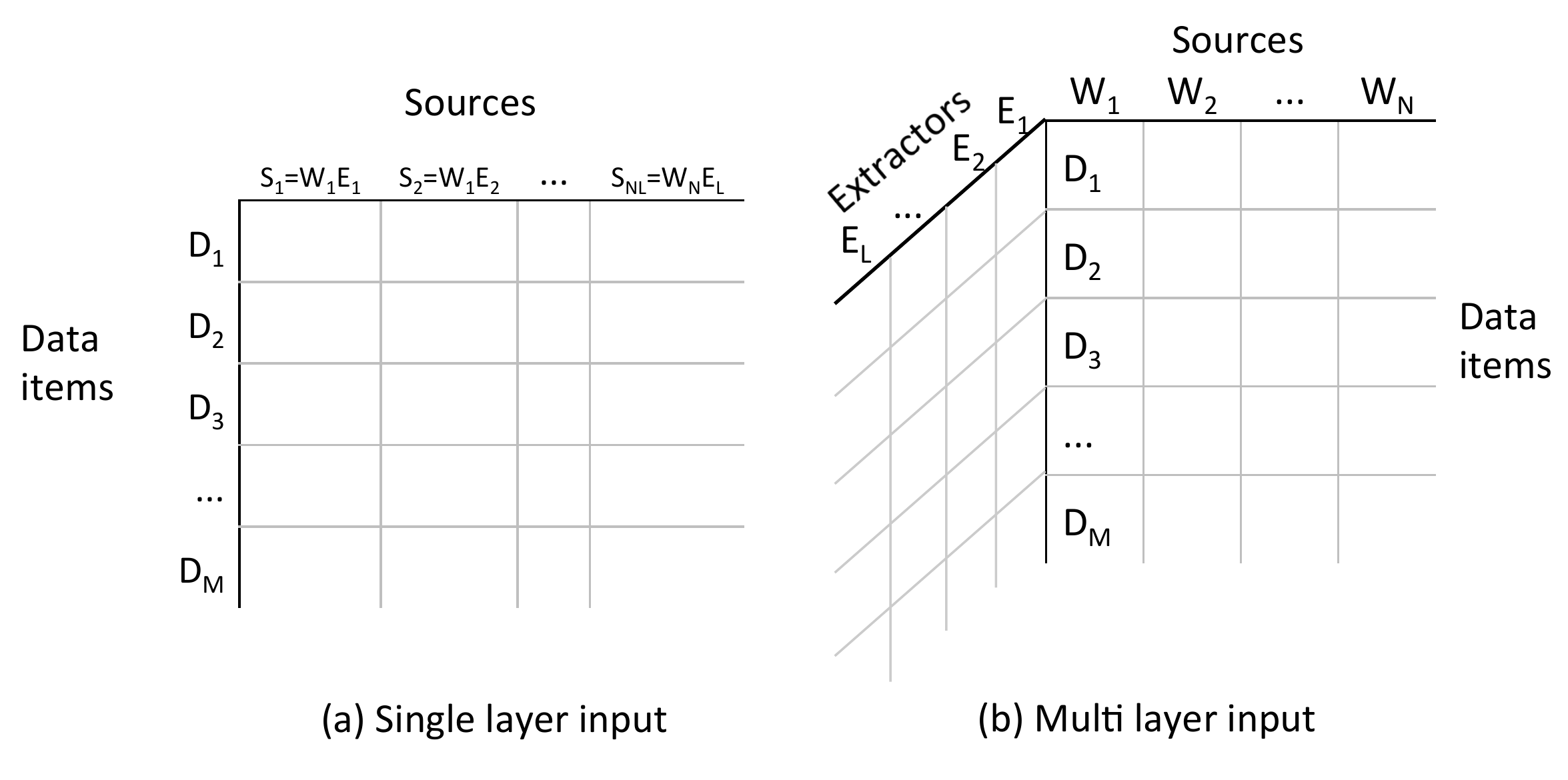}
\vspace{-.1in}
\caption{Form of the input data for (a) the single-layer model and (b)
  the multi-layer model.
\label{fig:kfcube}}
\vspace{-.2in}
\end{figure}

\subsection{Problem definition}
\label{sec:defn}
\noindent
We are given a set of web sources $\cal W$ and a set of extractors $\cal E$.
An extractor  is a method for extracting
{\tt (subject, predicate, object)} triples from a webpage.
For example, one extractor might look for the {\em pattern}
 {\em ``$\$A$, the president of $\$B$, ...''}, from which it can
 extract the triple {\em ($A$, nationality, $B$)}.
Of course, this is not always correct (\eg, if $A$ is the
president of a company, not a country).
In addition, an extractor reconciles the string representations of entities 
into entity identifiers such as Freebase mids,
and sometimes this fails too.
It is the presence of these common extractor errors,
which are separate from source errors (\ie, incorrect claims on a
webpage), that motivates our work.

In the rest of the paper, we represent such triples as (data item, value) pairs,
where the data item is in the form of {\tt (subject, predicate)},
describing a particular aspect of an entity, and the {\tt object} serves as a
value for the data item.
We summarize the notation used in this paper in Table~\ref{tbl:notation}.

We define an observation variable $X_{ewdv}$.
We set $X_{ewdv}=1$ if extractor $e$ extracted value
 $v$ for data item $d$ on web source $w$; if it did not extract such a
 value, we set $X_{ewdv}=0$. 
An extractor might also return confidence values indicating how confident
it is in the correctness of the extraction;
we consider these extensions in Section~\ref{sec:confidenceWeightedExtractions}.
We use 
matrix $X= \{X_{ewdv}\}$ to denote all the data.

We can represent $X$ as a (sparse) ``data cube'', as shown in
Figure~\ref{fig:kfcube}(b).
Table~\ref{tbl:data} shows an example of a single horizontal ``slice''
of this cube for the case where the data item is $d^*=$ {\em (Barack Obama,
nationality).}
We discuss this example in more detail next.

{
\begin{table}
\small
\vspace{-.1in}
\center
\caption{Obama's nationality extracted by 5 extractors from 8 webpages.
Column 2 (Value) shows the nationality truly
provided by each source; Columns 3-7 show the nationality extracted by
each extractor. Wrong extractions are shown in italics. \label{tbl:data}}
\vspace{-.1in}
\begin{tabular}{|c|c||c|c|c|c|c|}
\hline
 & Value & $E_1$ & $E_2$ & $E_3$ & $E_4$ & $E_5$ \\
\hline
$W_1$ & USA & USA & USA & USA & USA & {\em Kenya} \\
$W_2$ & USA & USA & USA & USA & {\em \Nam} & \\
$W_3$ & USA & USA & & USA & {\em N. Amer.} & \\
$W_4$ & USA & USA & & USA & {\em Kenya} & \\
$W_5$ & Kenya & Kenya & Kenya & Kenya & Kenya & Kenya \\
$W_6$ & Kenya & Kenya &  & Kenya & {\em USA} & \\
$W_7$ & - & & & {\em Kenya} & & {\em Kenya} \\
$W_8$ & - & & & & & {\em Kenya} \\
\hline
\end{tabular}
\vspace{-.1in}
\end{table}
}

\begin{example}
\label{eg:motivating}
\smallex
Suppose we have 8 webpages, $W_1-W_8$,
and suppose we are interested in the data item
 {\em (Obama, nationality)}.
The value stated for this data item by each of the webpages is shown in
the left hand column of Table~\ref{tbl:data}.
We see that $W_1-W_4$ provide {\em USA} as the nationality of
Obama, whereas $W_5-W_6$ provide {\em Kenya} (a false value). Pages $W_7-W_8$ do not
provide any information regarding Obama's nationality.

Now suppose we have 5 different extractors of varying reliability.
The values they extract for this data item from each of the 8 webpages 
are shown in the table.
Extractor $E_1$ extracts all the provided triples correctly.
Extractor $E_2$ misses some of the provided triples (false negatives), but all of its extractions are correct.
Extractor $E_3$ extracts all the provided triples, but also wrongly
extracts the value {\em Kenya} from $W_7$, even though $W_7$ does not
provide this value (a false positive).
Extractor $E_4$ and $E_5$ both have poor quality, missing a lot
of provided triples and making numerous mistakes.
%
\rbox
\end{example}

\noindent
For each web source $w \in \cal W$,
we define its {\em accuracy}, denoted by $A_w$, as the probability that 
a value it provides for a fact is correct (\ie, consistent with the real world).
We use $A=\{A_w\}$ for the set of all accuracy parameters. 
Finally, we can formally define the problem of KBT estimation.

\vspace{-.1in}
\begin{definition}[KBT Estimation]
The {\em 
Knowledge-Based Trust (KBT) estimation task} is to  estimate the web source accuracies
$A=\{A_w\}$ given the observation matrix $X=\{X_{ewdv}\}$ of extracted triples.
\rbox
\end{definition}

\subsection{Estimating the truth using a single-layer model}
\label{sec:singleLayer}
KBT estimation is closely related to the {\em knowledge fusion} problem we
studied in our previous work~\cite{DGE+14}, where we evaluate the true
(but latent) values for each of the data items, given the noisy observations. 
We introduce the binary latent variables $T_{dv}$, 
which represent whether $v$ is a correct value for data item $d$.
Let $T=\{T_{dv}\}$. {\em Given the observation matrix $X=\{X_{ewdv}\}$,
the knowledge fusion problem computes the posterior over the latent variables, $p(T|X)$.}

One way to solve this problem is to
``reshape'' the cube into a two-dimensional matrix, as shown in 
Figure~\ref{fig:kfcube}(a), by treating every combination of web
page and extractor as a distinct data source.
Now the data are in a form that standard {\em data fusion} techniques
(surveyed in~\cite{LDL+12}) expect.
We call this a {\em single-layer model}, since it only has one layer of
latent variables (representing the unknown values for the data items).
We now review this model in detail, and we compare it with our work shortly.

In our previous work \cite{DGE+14}, we applied the  probabilistic model
described in~\cite{DBS09a}.
We assume that each data item can only have a single true value.
This assumption holds for functional predicates, such as
{\em nationality} or {\em date-of-birth}, but is not technically valid for
set-valued predicates, such as {\em child}.
Nevertheless, \cite{DGE+14}  showed empirically that this ``single truth''
assumption works well in practice  even for non-functional predicates,
so we shall adopt it in this work for simplicity.
(See \cite{PDD+14, ZRHG12} for approaches to dealing with multi-valued attributes.)

Based on the single-truth assumption,
we define a latent variable $V_d \in \dom(d)$ for each data item
to present the true value for $d$,
where $\dom(d)$ is the domain (set of possible values) for data item $d$. 
Let $V=\{V_d\}$ and note that we can derive $T=\{T_{dv}\}$ from $V$ 
under the single-truth assumption.
We then define the following observation model:
\begin{equation}
p(X_{sdv}=1 | V_d=v^*, A_s)
  = \left\{ \begin{array}{ll}
    A_s & \mbox{ if } v=v^* \\
    \frac{1-A_s}{n} & \mbox{ if } v \neq v^* 
\end{array} \right.
\label{eqn:CPDD}
\end{equation}
where $v^*$ is the true value, $s=(w,e)$ is the source, $A_s \in [0,1]$ is the {\em accuracy}
of this data source, and $n$ is the number of false
values for this domain (\ie, we assume $|\dom(d)|=n+1$).
The model says that the probability for $s$ to provide a true value
$v^*$ for $d$ is its accuracy, whereas the probability for it
to provide one of the $n$ false values is $1-A_s$ divided by $n$.

\newcommand{\allA}{A}

Given this model, it is simple to apply Bayes rule to compute
$p(V_d|X_d, \allA)$, where $X_d=\{X_{sdv}\}$ is all the data pertaining to
data item $d$ (\ie, the $d$'th row of the data matrix),
and $\allA=\{A_s\}$ is the set of all accuracy parameters.
Assuming a uniform prior for $p(V_d)$,
this can be done as follows:
\begin{equation}
p(V_d=v|X_d,\allA) = \frac{p(X_d|V_d=v,\allA)}{\sum_{v' \in \dom(d)} p(X_d|V_d=v',\allA)}
\label{eqn:pVgivenD}
\end{equation}
where the likelihood function can be derived from Equation~(\ref{eqn:CPDD}),
assuming independence of the data sources:\footnote{\small Previous works~\cite{DBS09a,PDD+14}
discussed how to detect copying and correlations between sources in data fusion; however,
scaling them up to billions of web sources remains an open problem.}
\begin{equation}
p(X_d|V_d=v^*,\allA) =   \prod_{s,v: X_{sdv}=1} p(X_{sdv} = 1 | V_d=v^*, A_s)
\label{eqn:likD}
\end{equation}

This model is called the \accu  model~\cite{DBS09a}.
A slightly more advanced model, known as \popaccu,
removes the assumption that the wrong values are uniformly
distributed. Instead, it uses the empirical distribution of values in the observed
data. It was proved that the \popaccu model is monotonic;
that is, adding more sources would not reduce the quality of results~\cite{DSS13}.

In both \accu and \popaccu, it is necessary to jointly estimate the
hidden values $V=\{V_d\}$ and the accuracy parameters $\allA=\{A_s\}$. 
An iterative EM-like algorithm was proposed for performing
this as follows (\cite{DBS09a}):
\begin{itemize}\tightlist
\item Set the iteration counter $t=0$.
\item Initialize the parameters $A_s^t$ to some value (\eg, 0.8).
\item Estimate 
$p(V_d|X_d,\allA^{t})$
in parallel for all $d$
using Equation~(\ref{eqn:pVgivenD}) (this is like the E step).
From this we can
compute the most probable value, $\hat{V}_d = \argmax p(V_d|X_d,A^{t})$.
\item  Estimate 
$\hat{A}^{(t+1)}_s$
as follows:
\begin{equation}
\hat{A}_s^{t+1}
 = \frac{\sum_{d} \sum_v \ind{X_{sdv}=1} p(V_d=v|X_d,A^{t})}
{\sum_{d} \sum_v \ind{X_{sdv}=1}}
\label{eqn:MstepA}
\end{equation}
where $\ind{a=b}$ is 1 if $a=b$ and is 0 otherwise.
Intuitively this equation says that we estimate the accuracy of a
source by the average probability of the facts it extracts.
This equation is like the M step in EM.
\item We now return to the E step, and iterate until convergence.
\end{itemize}
Theoretical properties of this algorithm are discussed in \cite{DBS09a}.

\subsection{Estimating KBT using a multi-layer model}
\label{sec:overview}
Although estimating KBT is closely related to knowledge fusion, 
the single-layer model falls short in two aspects to solve the new problem. 
The first issue is its inability to assess trustworthiness
of web sources independently of extractors; in other words,
$A_s$ is the accuracy of a $(w,e)$ pair, rather than the accuracy of a web source itself.
Simply assuming all extracted values are actually provided by the source obviously would not work.
    In our example, we may wrongly infer that
    $W_1$ is a bad source because of the extracted {\em Kenya} value,
    although this is an extraction error. \eat{; we may also wrongly infer
    that $E_1$ is a bad extractor because 2 out of 6 of its extracted triples
    are incorrect, although these are actually erroneous values provided
    by sources $W_5-W_6$. }

The second issue is the inability to properly assess truthfulness of 
    triples. In our example,
    there are 12 sources (\ie, extractor-webpage pairs) for {\em USA} and 12 sources for {\em Kenya};
this seems to suggest
    that {\em USA} and {\em Kenya} are equally likely to be true.
    However, intuitively this seems unreasonable:
extractors $E_1-E_3$ all tend to agree with each other, and so seem to
be reliable; we can therefore ``explain away'' the Kenya values extracted
by $E_4-E_5$ as being more likely to be extraction errors.

Solving these two problems requires us to
distinguish extraction errors from source errors.
In our example, we wish to distinguish correctly extracted true triples
(\eg, {\em USA} from $W_1-W_4$), correctly extracted false triples
(\eg, {\em Kenya} from $W_5-W_6$), wrongly extracted true triples
(\eg, {\em USA} from $W_6$), and wrongly extracted false triples
(\eg, {\em Kenya} from $W_1, W_4, W_7-W_8$).

In this paper, we present a new probabilistic model that can 
estimate the accuracy of each web source, factoring out the noise
introduced by the extractors. It differs from the single-layer model 
in two ways. First, in addition to the latent variables to
represent the true value of each data item ($V_d$), the new model introduces a set
of latent variables to represent whether each extraction was correct
or not; this allows us to distinguish extraction errors and
source data errors. Second, instead of using $\allA$ to represent the accuracy
of $(e,w)$ pairs, the new model defines a set of parameters for the accuracy
of the web sources, and for the quality of the extractors;
this allows us to separate the quality of the sources from that of the extractors.
We call the new model the {\em multi-layer model}, because it contains 
two layers of latent variables and parameters (Section~\ref{sec:fusion}).

The fundamental differences between the multi-layer model and
the single-layer model allow for reliable KBT estimation. 
In Section~\ref{sec:hierarchy}, we also show how to 
dynamically select the granularity of a source and an extractor.
Finally, in Section~\ref{sec:exp}, 
we show empirically how both components play
an important role in improving the performance over the single-layer model.

\section{Multi-Layer Model}
\label{sec:fusion}
In this section, we describe in detail how we compute $A=\{A_w\}$
from our observation matrix $X=\{X_{ewdv}\}$ using a multi-layer model. 

\subsection{The multi-layer model}
\label{sec:multi-layer}

We extend the previous single-layer model in two ways.
First, we introduce the binary latent variables $C_{wdv}$, which
represent whether web source $w$ actually provides triple
$(d,v)$ or not. Similar to Equation~(\ref{eqn:CPDD}),
these variables depend on the true values $V_d$ and the accuracies of
each of the web sources $A_w$ as follows:
\begin{equation}
p(C_{wdv}=1 | V_d=v^*, A_w)
  = \left\{ \begin{array}{ll}
    A_w & \mbox{ if } v=v^* \\
    \frac{1-A_w}{n} & \mbox{ if } v \neq v^* 
\end{array} \right.
\label{eqn:CPDC}
\end{equation}

Second,  following \cite{PDD+14, ZRHG12}, we use a two-parameter noise
model for the observed data, as follows:
\begin{equation}
p(X_{ewdv}=1 | C_{wdv}=c, Q_e, R_e)
  = \left\{ \begin{array}{ll}
    R_e & \mbox{ if } c=1 \\
    Q_e  &\mbox{ if } c=0
\end{array} \right.
\label{eqn:CPDX}
\end{equation}
Here $R_e$ is the {\em recall} of the extractor; that is, the probability of extracting
a truly provided triple. And $Q_e$ is 1 minus the {\em specificity}; that is, the
probability of extracting an unprovided triple.
Parameter $Q_e$ is related to the recall ($R_e$) and precision ($P_e$) as follows:
\begin{equation}
Q_e = \frac{\gamma}{1-\gamma} \cdot \frac{1-P_e}{P_e} \cdot R_e
\label{eqn:Q}
\end{equation}
where $\gamma = p(C_{wdv}=1)$ for any $v \in \dom(d)$,
as explained in \cite{PDD+14}.
(Table~\ref{tbl:extractor} gives a numerical example of computing
$Q_e$ from $P_e$ and $R_e$.)

To complete the specification of the model, we must specify the prior probability of the various model parameters:
\begin{equation}
\theta_1=\{A_w\}_{w=1}^W, \theta_2 = (\{P_e\}_{e=1}^E, \{R_e\}_{e=1}^E), \theta = (\theta_1, \theta_2)
\end{equation}
For simplicity, we use uniform priors on the parameters.
By default,  we set $A_w=0.8$, $R_e=0.8$, and $Q_e=0.2$.
In Section~\ref{sec:exp}, we discuss an alternative way to
estimate the initial value of $A_w$, based on the fraction of correct
triples that have been extracted from this source, using an external
estimate of correctness (based on {\em Freebase}~\cite{freebase}).

\begin{figure}
\vspace{-.1in}
\center
\includegraphics[height=2in]{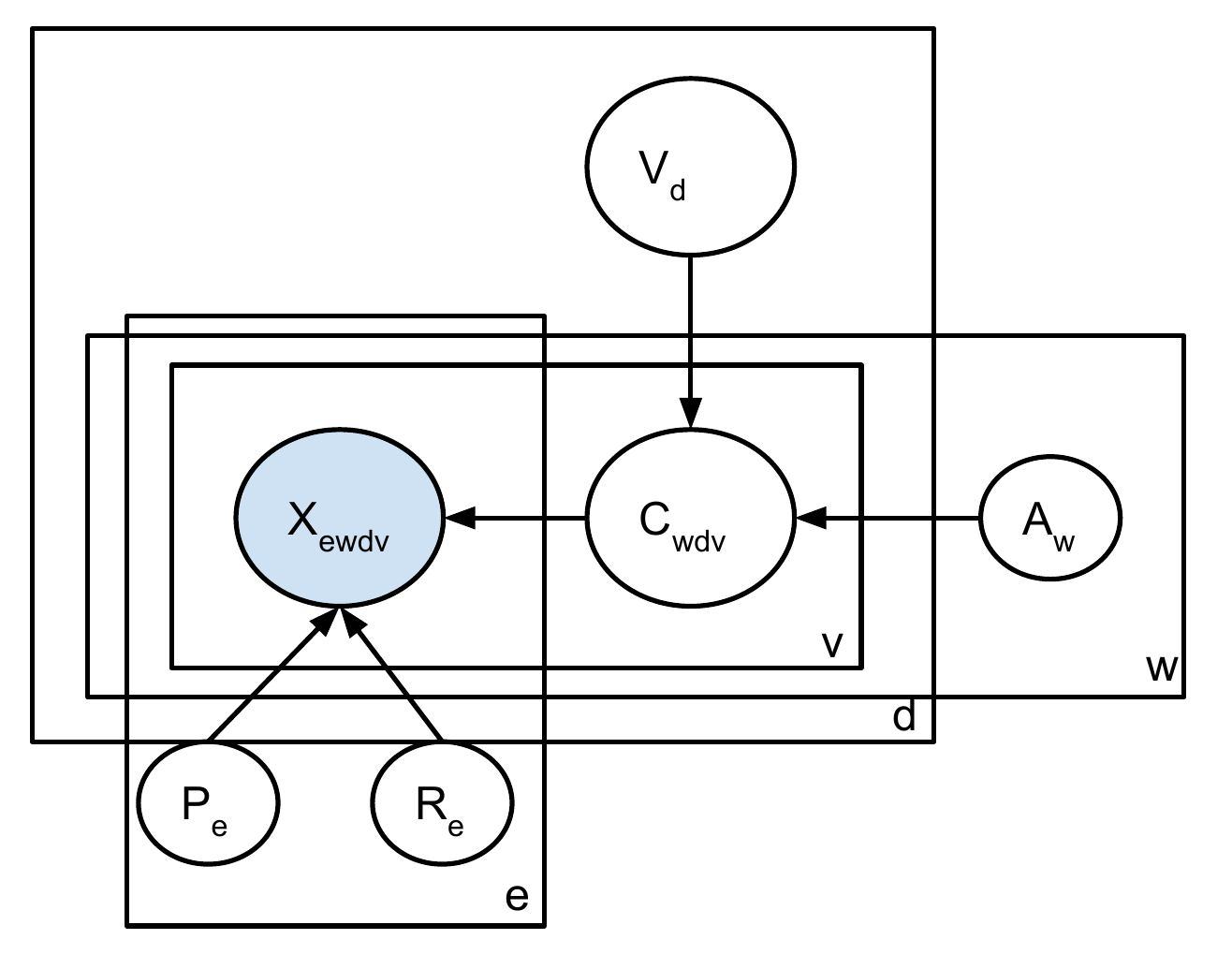}
\vspace{-.1in}
\caption{A representation of the multi-layer model using graphical model plate notation.}
\label{fig:gmPlates}
\end{figure}

Let $V=\{V_d\}$, $C= \{C_{wdv}\}$, and $Z=(V,C)$ be all the latent variables.
Our model defines the following joint distribution:
\begin{equation}
p(X, Z, \theta) 
 = p(\theta) p(V) p(C|V,\theta_1) p(X|C,\theta_2)
\end{equation}
We can represent the conditional independence assumptions we are
making using a graphical model,
as shown in Figure~\ref{fig:gmPlates}.
The shaded node is an observed variable, representing the data;
the unshaded nodes are hidden variables or parameters.
The arrows indicate the dependence between the variables and parameters.
The boxes are known as ``plates'' and represent repetition of the enclosed variables;
for example, the box of $e$ repeats for every extractor $e \in \cal E$.
\eat{To make this clearer, in Figure~\ref{fig:GMex} we have partially
``unrolled'' the model, showing that for each $(d,v)$ slice
of the four-dimensional $X$ tensor, we have a two-dimensional matrix, where each row 
is associated with a source  accuracy parameter $A_w$,
and each column is associated with two extractor quality parameters
$P_e$ and $R_e$.} 

\eat{
\begin{figure}
\center
\includegraphics[height=2in]{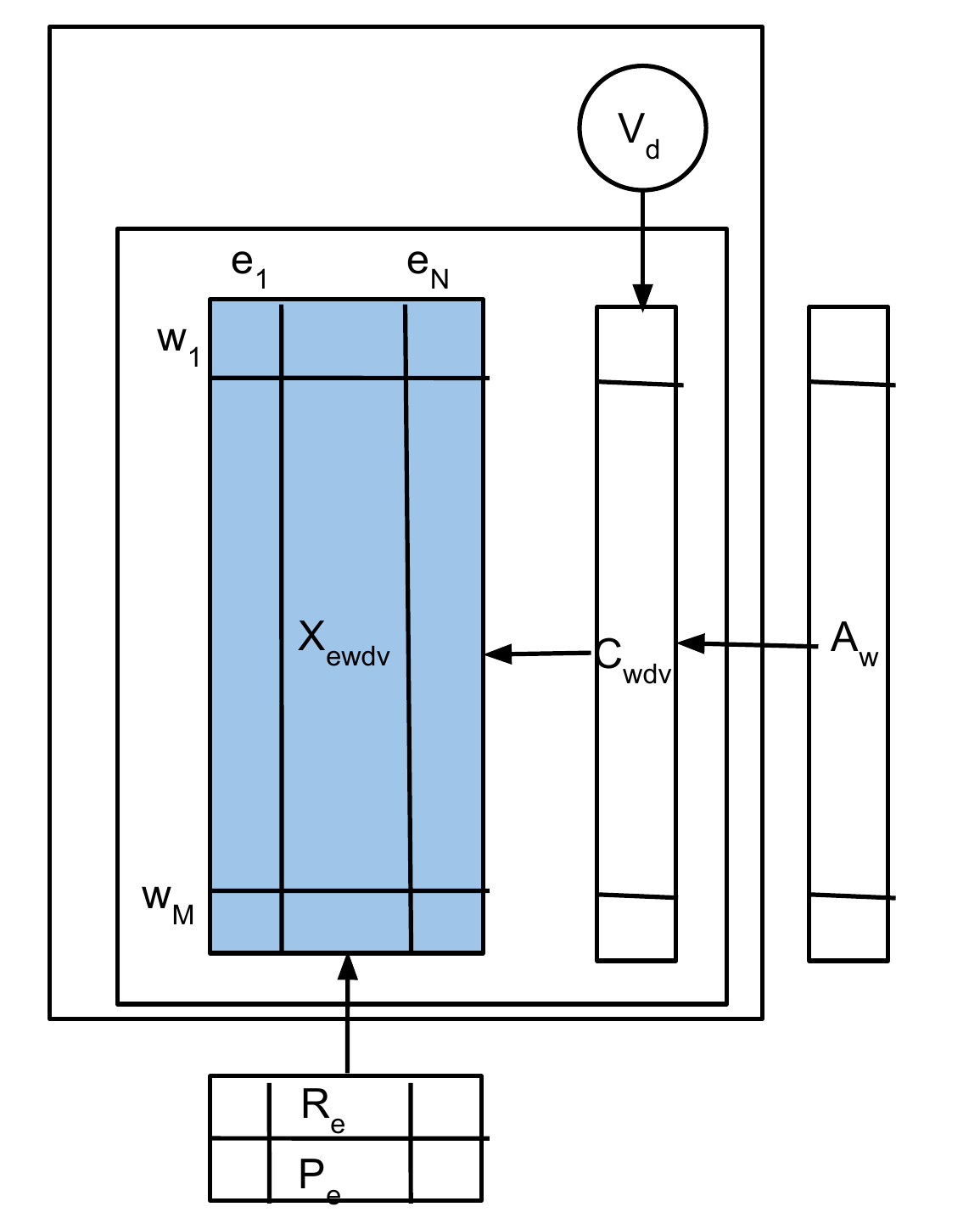}
\caption{An example of the graphical model 
where we have ``unrolled'' the plate over the $M$ web sources and the
$N$ extractors.
}
\label{fig:gmEx}
\label{fig:GMex}
\vspace{-.1in}
\end{figure}
}

{\small 
\begin{algorithm}[t]
\caption{\sc{MultiLayer}($X, t_{max}$)
\label{algo:multiLayer}}
\Input{$X$: all extracted data; \\
$t_{max}$: max number of iterations.}
\Output{Estimates of $Z$ and $\theta$.}
        Initialize $\theta$ to default values; \\
          \For { $t \in [1,t_{max}]$ } {
            Estimate $C$ by Eqs.(\ref{eqn:multiTruth2}, \ref{eqn:prior}, \ref{eqn:voteCountCSoft});\\
            Estimate $V$ by Eqs.(\ref{eqn:voteCountValueSoft1}-\ref{eqn:voteCountValueSoft2});\\
            Estimate $\theta_1$ by Eq.(\ref{eqn:prdata});\\
            Estimate $\theta_2$ by Eqs.(\ref{eqn:pr1}-\ref{eqn:pr2});\\
            \If {$Z, \theta$ converge} {
              {\bf break};
            }
          }

	\Return $Z, \theta$;
\end{algorithm}
}
\subsection{Inference}
\label{sec:inference}
Recall that estimating KBT essentially requires us
to compute the posterior over the parameters of
interest, $p(A | X)$. Doing this exactly is computationally
intractable, because of the presence of the latent variables $Z$.
One approach is to use a Monte Carlo approximation,
such as Gibbs sampling,
as in \cite{ZH12}. However, this can be slow and is hard to
implement in a Map-Reduce framework, which is required for the scale
of data we use in this paper.

A faster alternative is to use EM, which will return a point estimate
of all the parameters, $\hat{\theta} = \argmax p(\theta|X)$.
Since we are using a uniform prior,
this  is equivalent to the maximum likelihood estimate
$\hat{\theta} = \argmax p(X | \theta)$.
From this, we can derive $\hat{A}$.

As pointed out in~\cite{PR13}, an exact EM algorithm has a quadratic 
complexity even for a single-layer model,
so is unaffordable for data of web scale.
Instead, we use an iterative ``EM like'' estimation procedure,
where we initialize the parameters as described previously, and then
alternate between estimating $Z$ and then estimating $\theta$,
until we converge.

We first given an overview of this EM-like algorithm, and then go into
details in the following sections.

In our case, $Z$ consists of two ``layers'' of variables. We update
them sequentially, as follows.
First, let $X_{wdv}=\{X_{ewdv}\}$ denote all extractions from web source
$w$ about a particular triple $t=(d,v)$.
We compute the extraction correctness $p(C_{wdv}|X_{wdv},\theta^t_2)$,
as explained in Section~\ref{sec:estC}, and then we compute
$\hat{C}_{wdv}=\argmax \\p(C_{wdv}|X_{wdv},\theta^t_2)$,
which is our best guess about the ``true contents'' of each web source.
This can be done in parallel over $d,w,v$.

Let $\hat{C}_d=\hat{C}_{wdv}$ denote all the estimated values 
for $d$ across the different websites.
We compute $p(V_d|\hat{C}_{d},\theta^t_1)$, 
as explained in Section~\ref{sec:estV},
and then we compute $\hat{V}_d = \argmax p(V_d|\hat{C}_{d}, \theta^t_1)$,
which is our best guess about the
``true value'' of each data item.
This can be done in parallel over $d$.

Having estimated the latent variables, we then estimate
$\theta^{t+1}$. This parameter update
also consists of two steps (but can be done
in parallel): estimating the source accuracies $\{A_w\}$ and the extractor
reliabilities $\{P_e, R_e\}$,
as explained in Section~\ref{sec:estQuality}.

Algorithm~\ref{algo:multiLayer} gives a summary of the pseudo code; we give the
details next.

\subsection{Estimating the latent variables}

We now give the details of how we estimate the latent variables $Z$.
For notational brevity, we drop the conditioning on $\theta^t$,
except where needed.

\subsubsection{Estimating extraction correctness}
\label{sec:extractionCorrectness}
\label{sec:estC}

We first describe how to compute $p(C_{wdv}=1 | X_{wdv})$,
following the ``multi-truth'' model of \cite{PDD+14}.
We will denote the prior probability
$p(C_{wdv}=1)$ by $\alpha$. In initial iterations, we initialize this to
$\alpha=0.5$. Note that by using a fixed prior,
we break the connection between $C_{wdv}$ and $V_d$ in the graphical model, as shown in Figure~\ref{fig:gmPlates}.
Thus, in subsequent iterations, we re-estimate $p(C_{wdv}=1)$
using the results of $V_d$ obtained from the previous iteration, 
as explained in Section~\ref{sec:reEstimatePrior}.

We use Bayes rule as follows:
\begin{eqnarray}
\lefteqn{p(C_{wdv}=1 | X_{wdv})} \nonumber\\
 &=& \frac{ \alpha p( X_{wdv} | C_{wdv}=1)}
{ \alpha p( X_{wdv} | C_{wdv}=1) + (1-\alpha) p( X_{wdv} | C_{wdv}=0)} \nonumber \\
&=& \frac{1}{
1 + \frac{1}{\frac{p( X_{wdv} | C_{wdv}=1)}{p( X_{wdv} | C_{wdv}=0) }
                    \frac{\alpha}{1-\alpha}
                   }
   } \nonumber \\
&=& \sigma\left( \log \frac{p( X_{wdv} | C_{wdv}=1)}{p( X_{wdv} | C_{wdv}=0) }
                    + \log \frac{\alpha}{1-\alpha} \right)
\label{eqn:multiTruth}
\end{eqnarray}
where $\sigma(x) \defeq \frac{1}{1+e^{-x}}$ is the {\em sigmoid} function.

Assuming independence of the extractors, 
and using Equation~(\ref{eqn:CPDX}), we can compute the likelihood ratio as follows:
\begin{equation}
\frac{p(X_{wdv} | C_{wdv}=1)}{p(X_{wdv} | C_{wdv}=0)}
 =
\prod_{e: X_{ewdv}=1} \frac{R_e}{Q_e}
\prod_{e: X_{ewdv}=0} \frac{1-R_e}{1-Q_e}
\label{eqn:logLikRatio}
\end{equation}

In other words, for each extractor we can compute a {\em presence vote}
$\pre_e$ for a triple that it extracts, and an {\em absence vote} of $\abs_e$
for a triple that it does not extract:
\begin{eqnarray}
\pre_e & \defeq & \log R_e- \log Q_e \\
\abs_e &\defeq& \log(1-R_e)- \log(1-Q_e).
\end{eqnarray}

For each triple $(w,d,v)$ we can compute 
its {\em vote count} as the sum of the presence votes
and the absence votes: 
\begin{equation}
VCC(w,d,v) \defeq \sum_{e: X_{ewdv}=1} \pre_e
 + \sum_{e: X_{ewdv}=0} \abs_e
\label{eqn:voteCountC}
\end{equation}

Accordingly, we can rewrite Equation~(\ref{eqn:multiTruth}) as follows.
\begin{equation}
\label{eqn:multiTruth2}
p(C_{wdv}=1 | X_{wdv})
 = \sigma\left( VCC(w,d,v) + \log \frac{\alpha}{1-\alpha} \right).
\end{equation}

{\small
\begin{table}
\center
\vspace{-.1in}
\caption{Quality and vote counts of extractors in the motivating example. 
We assume $\gamma=.25$ when we derive $Q_e$ from $P_e$ and $R_e$.
\label{tbl:extractor}}
\vspace{-.2in}
\begin{tabular}{|c|c|c|c|c|c|}
\hline
 & $E_1$ & $E_2$ & $E_3$ & $E_4$ & $E_5$ \\
\hline
\hline
$Q(E_i)$ & .01 & .01 & .06 & .22 & .17 \\
$R(E_i)$ & .99 & .5  & .99 & .33 & .17 \\
$P(E_i)$ & .99 & .99 & .85 & .33 & .25 \\
\hline
$Pre(E_i)$ & 4.6 & 3.9 & 2.8 & .4 & 0 \\
$Abs(E_i)$ & -4.6 & -.7 & -4.5 & -.15 & 0 \\
\hline
\end{tabular}
\vspace{-.15in}
\end{table}
}

{\small
\begin{table}
\center
\vspace{-.1in}
\caption{Extraction correctness and data item value distribution 
  for the data in Table~\ref{tbl:data},
 using the extraction parameters in Table~\ref{tbl:extractor}.
Columns 2-4 show $p(C_{wdv}=1|X_{wdv})$,
as explained in Example~\ref{eg:multiTruth}.
The last row shows $p(V_{d}|\hat{C}_{d})$,
as explained in Example~\ref{eg:singleTruth2};
note that this distribution does not sum to 1.0,
since not all of the values are shown in the table.
\label{tbl:results}}
\vspace{-.1in}
\begin{tabular}{|c||c|c|c|}
\hline
 & USA & Kenya & N.Amer. \\
\hline
$W_1$ & 1 & 0 & - \\
$W_2$ & 1 & - & 0 \\
$W_3$ & 1 & - & 0 \\
$W_4$ & 1 & 0 & - \\
$W_5$ & - & 1 & - \\
$W_6$ & 0 & 1 & - \\
$W_7$ & - & .07 & - \\
$W_8$ & - & 0 & - \\
\hline
$p(V_d|\hat{C}_d)$ & .995 & .004 & 0 \\
\hline
\end{tabular}
\vspace{-.15in}
\end{table}
}

\begin{example}\label{eg:multiTruth}
\smallex
Consider the extractors in the motivating example (Table~\ref{tbl:data}).
Suppose we know $Q_e$ and $R_e$ for each extractor $e$
as shown in Table~\ref{tbl:extractor}. We can then
compute $\pre_e$ and $\abs_e$ as shown in the same table. 
We observe that in general, an extractor with low $Q_e$
(unlikely to extract an unprovided triple; \eg, $E_1, E_2$)
often has a high presence vote;
an extractor with high $R_e$ (likely to extract a provided triple; 
\eg, $E_1, E_3$) often has a low (negative) absence 
vote; and a low-quality extractor
(\eg, $E_5$) often has a low presence vote and a high absence vote.

Now consider applying Equation~\ref{eqn:multiTruth2} to compute the likelihood 
that a particular source provides the triple $t^*=${\em (Obama,
  nationality, USA)},
assuming $\alpha=0.5$.
For source $W_1$, we see that extractors $E_1-E_4$ extract $t^*$,
so the vote count is $(4.6+3.9+2.8+0.4)+(0)=11.7$
and hence $p(C_{1,t^*}=1|X_{w,t^*}) =\sigma(11.7)=1$.
For source $W_6$, we see that only $E_4$ extracts $t^*$,
so  the vote count is $(0.4)+(-4.6 -0.7 -4.5 -0)=-9.4$,
and hence $p(C_{6,t^*}=1 | X_{6,t^*}))=\sigma(-9.4)=0$. 
Some other values for $P(C_{wt}=1|X_{wt})$ are shown in
Table~\ref{tbl:results}.
\rbox
\end{example}

Having computed $p(C_{wdv}=1|X_{wdv})$, we can compute
$\hat{C}_{wdv} = \argmax p(C_{wdv}|X_{wdv})$.
This serves as the input to the next step of inference.

\subsubsection{Estimating true value of the data item}
\label{sec:estV}

In this step, we compute $p(V_d=v|\hat{C}_d)$,
following the ``single truth'' model of
\cite{DBS09a}.
By Bayes rule we have
\begin{equation}
p(V_d=v|\hat{C}_d)
 = \frac{p(\hat{C}_d | V_d=v) p(V_d=v)}
{\sum_{v' \in \dom(d)} p(\hat{C}_d | V_d=v') p(V_d=v')}
\label{eqn:singleTruth}
\end{equation}
Since we do not assume any prior knowledge of the
correct values, we assume a uniform prior $p(V_d=v)$, so we just need to focus on the likelihood.
Using Equation~(\ref{eqn:CPDC}), we have
\begin{eqnarray}
\lefteqn{p(\hat{C}_d | V_d=v)} \nonumber \\
 &=& 
\prod_{w: \hat{C}_{wdv}=1} A_w
\prod_{w: \hat{C}_{wdv}=0} \frac{1-A_w}{n}
\\
 &=& 
\prod_{w: \hat{C}_{wdv}=1} \frac{n A_w}{1-A_w}
\prod_{w: \hat{C}_{wdv} \in \{0,1\}} \frac{1-A_w}{n}
\end{eqnarray}
Since the latter term $\prod_{w: \hat{C}_{wdv} \in \{0,1\}} \frac{1-A_w}{n} $
is constant with respect to $v$, we can drop it. 

Now let us define the vote count as follows:
\begin{equation}
VCV(w)  \defeq \log \frac{n A_w}{1-A_w}
\end{equation}
Aggregating over web sources that provide this triple, we define
\begin{equation}
VCV(d,v)  \defeq \sum_{w} \ind{\hat{C}_{wdv}=1} VCV(w)
\label{eqn:voteCountValue}
\end{equation}
With this notation, we can rewrite Equation~(\ref{eqn:singleTruth}) as
\begin{equation}
\label{eqn:singleTruth2}
p(V_d=v|\hat{C}_d)
 = \frac{\exp(VCV(d,v))}
{\sum_{v' \in \dom(d)} \exp(VCV(d,v'))}
\end{equation}

\begin{example}\label{eg:singleTruth2}
\smallex
Assume we have correctly decided the triple provided by each web source,
as in the ``Value'' column of Table~\ref{tbl:data}.
Assume each source has the same accuracy $A_w=0.6$ and $n=10$, 
so the vote count is $\ln({10 * 0.6 \over 1-0.6})=2.7$.
Then {\em USA} has vote count $2.7*4=10.8$, {\em Kenya} has vote count
$2.7*2=5.4$, and an unprovided value, such as {\em NAmer},
has vote count $0$. Since there are 10 false values in the domain,
so there are 9 unprovided values.
Hence we have
$p(V_d=USA|\hat{C}_d) = \frac{\exp(10.8)}{Z}=0.995$,
where $Z=\exp(10.8) + \exp(5.4) + \exp(0)*9$.
Similarly,
$p(V_d=Kenya|\hat{C}_d) = \frac{exp(5.4)}{Z} = 0.004$.
This is shown in the last row of Table~\ref{tbl:results}.
The missing mass of $1-(0.995+0.004)$ is
assigned (uniformly) to the other 9 values that were not observed
(but in the domain).
\end{example}


\subsubsection{An improved estimation procedure}
\label{sec:improved}

So far, we have assumed that we first compute a MAP estimate
$\hat{C}_{wdv}$, which we then use as evidence for estimating $V_d$.
However, this ignores the uncertainty in $\hat{C}$.
The correct thing to do is to compute $p(V_d|X_d)$ marginalizing out
over $C_{wdv}$.
\begin{eqnarray}
p(V_d|X_d)
&\propto& P(V_d) P(X_d | V_d) \nonumber \\
 &=& 
p(V_d) \sum_{\vec{c}}  p(C_d=\vec{c} |V_d) p(X_d|C_d) 
\end{eqnarray}
Here we can consider each $\vec{c}$ as a {\em possible world},
where each element $c_{wdv}$ indicates whether
a source $w$ provides a triple $(d,v)$ (value 1) or not (value 0).

As a simple heuristic approximation to this approach,
we replace the previous vote counting
with a weighted version, as follows:
\begin{eqnarray}
\label{eqn:voteCountValueSoft1}
VCV'(w,d,v)  &\defeq& p(C_{wdv}=1 | X_d) \log \frac{n A_w}{1-A_w}  \\
VCV'(d,v)  &\defeq& \sum_{w} VCV'(d,w,v)
\end{eqnarray}
We then compute
\begin{equation}
p(V_d=v|X_d) \approx
  \frac{\exp(VCV'(d,v))}
{\sum_{v' \in \dom(d)} \exp(VCV'(d,v'))}
\label{eqn:voteCountValueSoft2}
\end{equation}
We will show that such improved estimation procedure
improves upon ignoring the uncertainty in $\hat{C}_d$
in experiments (Section~\ref{sec:exp_alt}).

\subsubsection{Re-estimating the prior of correctness}
\label{sec:topDown}
\label{sec:reEstimatePrior}

In Section~\ref{sec:extractionCorrectness},
we assumed that $p(C_{wdv}=1) = \alpha$ was known, which breaks
the connection between $V_d$ and $C_{wdv}$.
Thus, we update this prior after each
iteration according to the correctness of the value and
the accuracy of the source:
\begin{equation}
\label{eqn:prior}
\hat{\alpha}^{t+1} = 
 p(V_d=v|X) A_w + (1-p(V_d=v|X))(1-A_w)
\end{equation}
We can then use this refined estimate in the following iteration.
We give an example of this process.

\begin{example}
\label{ex:prior}
\smallex
Consider the probability that $W_7$ provides $t'=$ (Obama, nationality, Kenya).
Two extractors extract $t'$ from $W_7$ and the vote count is -2.65,
so the initial estimate is $p(C_{wdv}=1|X)=\sigma(-2.65)=0.06$.
However, after the previous iteration has finished,
we know that $p(V_d=Kenya|X) = 0.04$.
This gives us a modified prior probability as follows:
$p'(C_{wt}=1) =0.004*0.6 + (1-0.004) *(1-0.6)=0.4$,
assuming $A_w=0.6$.
Hence the updated posterior probability is given by
$p'(C_{wt}=1|X) = \sigma(-2.65 + \log {1-0.4 \over 0.4}) = 0.04$,
which is lower than before.
\end{example}

\subsection{Estimating the quality parameters}
\label{sec:estQuality}

Having estimated the latent variables, we now estimate the parameters of the model.

\subsubsection{Source quality}

Following \cite{DBS09a}, we estimate the accuracy of a source by computing the average probability of its provided values being true:
\begin{equation}
\hat{A}_w^{t+1}
 = \frac{\sum_{dv: \hat{C}_{wdv}=1} p(V_d=v|X)}
{\sum_{dv: \hat{C}_{wdv}=1} 1 }
\end{equation}
We can take uncertainty of $\hat{C}$ into account as follows:
\begin{equation}
\hat{A}_w^{t+1}
 = \frac{\sum_{dv: \hat{C}_{wdv}>0} p(C_{wdv}=1 | X) p(V_d=v|X)}
{\sum_{dv: \hat{C}_{wdv}>0} p(C_{wdv}=1|X)}
\label{eqn:prdata}
\end{equation}
{\em This is the key equation behind Knowledge-based Trust estimation}:
it estimates the accuracy of a web source as the weighted average of
the probability of the facts that it contains (provides), where the weights are
the probability that these facts are indeed contained in that source.

\subsubsection{Extractor quality}
According to the definition of precision and recall,
we can estimate them as follows:
\begin{eqnarray}
\hat{P}_e^{t+1}
 &=& \frac{\sum_{wdv: X_{ewdv}=1} p(C_{wdv}=1|X)}
{\sum_{wdv: X_{ewdv}=1} 1} \\
\hat{R}_e^{t+1}
 &=& \frac{\sum_{wdv: X_{ewdv}=1} p(C_{wdv}=1|X)}
{\sum_{wdv} p(C_{wdv}=1|X) }
\end{eqnarray}
Note that for reasons explained in \cite{PDD+14},
it is much more reliable to estimate $P_e$ and $R_e$ from data,
and then compute $Q_e$ using Equation~(\ref{eqn:Q}), rather than trying
to estimate $Q_e$ directly.

\subsection{Handling confidence-weighted extractions}
\label{sec:confidenceWeightedExtractions}

So far, we have assumed that each extractor returns a binary decision about whether it extracts a triple or not,
$X_{ewdv} \in \{0,1\}$. However, in real life, extractors return confidence scores, which we can interpret as the
probability that the triple is present on the page according to that
extractor. Let us denote this ``soft evidence'' by $p(X_{ewdv}=1)  =
\overline{X}_{ewdv} \in [0,1]$. A simple way to handle such data is to
binarize it, by thresholding. However, this loses information, as
shown in the following example. 

\begin{example}\label{eg:prData}
\smallex
Consider the case that $E_1$ and $E_3$ are not fully confident 
with their extractions from $W_3$ and $W_4$. In particular,
$E_1$ gives each extraction a probability (\ie, confidence) .85, 
and $E_3$ gives probability .5.
Although no extractor has full confidence for the extraction, 
after observing their extractions collectively,
we would be fairly confident that $W_3$ and $W_4$
indeed provide triple $T=${\em (Obama, nationality, USA)}. 

However, if we simply apply a threshold of .7, we would ignore the extractions 
from $W_3$ and $W_4$ by $E_3$. Because of lack of extraction,
we would conclude that neither $W_3$ nor $W_4$ provides $T$.
Then, since {\em USA} is provided by $W_1$ and $W_2$, whereas {\em Kenya}
is provided by $W_5$ and $W_6$, and the sources all have the same accuracy,
we would compute an equal probability for {\em USA} and for {\em Kenya}. \rbox
\end{example}

Following the same approach as in Equation~(\ref{eqn:voteCountValueSoft1}), we propose
to modify Equation~(\ref{eqn:voteCountC})
 as follows:
\begin{equation}
VCC'(w,d,v) \defeq \sum_e \left[
p(X_{ewt}=1) \pre_e + p(X_{ewt}=0) \abs_e
\right]
\label{eqn:voteCountCSoft}
\end{equation}

Similarly, we modify the precision and recall estimates:
\begin{eqnarray}
\label{eqn:pr1}
\hat{P}_e
 &=& \frac{\sum_{wdv: \overline{X}_{ewdv}>0} p(X_{ewdv}=1) p(C_{wdv}=1|X)}
{\sum_{wdv: \overline{X}_{ewdv}>0} p(X_{ewdv}=1)} \\
\hat{R}_e
 &=& \frac{\sum_{wdv: \overline{X}_{ewdv}>0} p(X_{ewdv}=1) p(C_{wdv}=1|X)}
{\sum_{wdv} p(C_{wdv}=1|X) }
\label{eqn:pr2}
\end{eqnarray}

\section{Dynamically Selecting Granularity}
\label{sec:hierarchy}
This section describes the choice of the granularity for web sources;
at the end of this section we discuss how to apply it to extractors. 
This step is conducted before applying the multi-layer model.

Ideally, we wish to use the finest granularity. For example,
it is natural to treat each webpage as a separate source, as it may 
have a different accuracy from other webpages. 
We may even define a source as a specific predicate
on a specific webpage; this allows us to estimate how trustworthy a
page is about a specific kind of predicate.
However, when we define sources too finely,
we may have too little data to reliably estimate their accuracies;
conversely, there may exist sources that have too much data 
even at their finest granularity, which can cause computational bottlenecks.

To handle this, we wish to dynamically choose the granularity 
of the sources. For too small sources, we can ``back off'' to a
coarser level of the hierarchy; this allows us to ``borrow statistical
strength'' between related pages. For too large sources, we may
choose to split it into multiple sources and estimate their accuracies
independently. When we do merging, our goal is to improve the statistical 
quality of our estimates without sacrificing efficiency.
When we do splitting, our goal is to significantly improve efficiency 
in presence of data skew, without changing our estimates dramatically.

To be more precise, 
we can define a source at multiple levels of resolution
by specifying the following values of a feature vector:
$\langle${\tt website, predicate, webpage}$\rangle$,
ordered from most general to most specific.
We can then arrange these sources in a hierarchy.
For example, $\langle${\em wiki.com}$\rangle$ 
is a parent of $\langle${\em wiki.com}, {\sf date\_of\_birth}$\rangle$, 
which in turn is a parent of 
$\langle${\em wiki.com}, {\sf date\_of\_birth}, {\em wiki.com/page1.html}$\rangle$.
We define the following two operators.

\begin{itemize}\tightlist
\item {\bf Split:} When we split a large source, we wish to split it randomly into sub-sources 
of similar sizes. Specifically,
let $W$ be a source with size $|W|$, and $M$ be the maximum size we desire;
we uniformly distribute the triples from $W$ into $\lceil {|W| \over M} \rceil$
buckets, each representing a sub-source. 
We set $M$ to a large number that does not require splitting sources
unnecessarily and meanwhile would not cause computational
bottleneck according to the system performance.

\item  {\bf Merge:} 
When we merge small sources, we wish to merge only sources that share 
some common features, 
such as sharing the same predicate, or coming from
the same website; hence we only merge children with the same parent
in the hierarchy.
We set $m$ to a small number that does not require merging sources
unnecessarily while maintaining enough statistical strength.
\end{itemize}

\begin{example}\label{eg:merge}
\smallex
Consider three sources: $\langle${\em website1.com}, \\{\sf date\_of\_birth}$\rangle$,
$\langle${\em website1.com}, {\sf place\_of\_birth}$\rangle$,
$\langle${\em website1.com}, \\{\sf gender}$\rangle$,
each with two triples, arguably not enough for quality evaluation.
We can merge them into their parent source by removing the second feature.
We then obtain a source $\langle${\em website1.com}$\rangle$ with size
$2*3=6$, which gives more data for quality evaluation. \rbox
\end{example}

Note that when we merge small sources, the result parent source may not be of
desired size: it may still be too small, or it may be too large after we merge 
a huge number of small sources. As a result, we might need to iteratively 
merge the resulting sources to their parents, or splitting an oversized 
resulting source, as we describe in the full algorithm.

{\small 
\begin{algorithm}[t]
\caption{\rm{SplitAndMerge}(${\bf W}, m, M$)
\label{algo:splitAndMerge}}
\Input{${\bf W}$: sources with finest granularity; \\
$m/M$: min/max source size in desire.}
\Output{${\bf W}'$: a new set of sources with desired size.}

        ${\bf W}' \leftarrow \emptyset$;\\ \label{ln:init}

          \For { $W \in {\bf W}$ } { \label{ln:start}
            ${\bf W} \leftarrow {\bf W} \setminus \{W\}$;\\
            \If {$|W| > M$} {
              ${\bf W}' \leftarrow {\bf W}' \cup$ {\sc Split}$(W)$; \label{ln:split}
            }\ElseIf {$|W| < m$} {
              $W_{par} \leftarrow $ {\sc GetParent} $(W)$; \label{ln:merge1}\\ 
              \If {$W_{par} = \perp$} { {\em // Already reach the top of the hierarchy} 
                ${\bf W}' \leftarrow {\bf W}' \cup \{W\}$; \label{ln:merge2}
              } \Else {
                ${\bf W} \leftarrow {\bf W} \cup \{W_{par}\}$; \label{ln:merge3}
              }
            }\Else {
              ${\bf W}' \leftarrow {\bf W}' \cup \{W\}$; \label{ln:remain}
            }
          }

	\Return ${\bf W}'$;
\end{algorithm}
}

Algorithm~\ref{algo:splitAndMerge} gives the {\sc SplitAndMerge} algorithm.
We use $\bf W$ for sources for examination and $\bf W'$ for final results;
at the beginning $\bf W$ contains all sources of the finest granularity and ${\bf W}'=\emptyset$
(Ln~\ref{ln:init}). We consider each $W \in {\bf W}$ (Ln~\ref{ln:start}). 
If $W$ is too large, we apply {\sc Split} to split it into a set of sub-sources;
{\sc Split} guarantees that each sub-source would be of desired size,
so we add the sub-sources to $\bf W'$ (Ln~\ref{ln:split}). If $W$ is too small, 
we obtain its parent source (Ln~\ref{ln:merge1}). In case $W$
is already at the top of the source hierarchy so it has no parent,
we add it to $\bf W'$ (Ln~\ref{ln:merge2}); otherwise, we add $W_{par}$ 
back to $\bf W$ (Ln~\ref{ln:merge3}).
Finally, for sources already in desired size, we move them directly to
$\bf W'$ (Ln~\ref{ln:remain}). 

\begin{example}\label{ex:splitAndMerge}
\smallex
Consider a set of 1000 sources $\langle W, P_i, URL_i \rangle, \\i \in [1,1000]$;
in other words, they belong to the same website, each has a different predicate
and a different URL. Assuming we wish to have sources with size in $[5, 500]$,
\SplitAndMerge proceeds in three stages.

In the first stage, each source is deemed too small and is replaced with
its parent source $\langle W, P_i \rangle$. In the second stage, each new source
is still deemed too small and is replaced with its parent source $\langle W \rangle$.
In the third stage, the single remaining source is deemed too large and
is split uniformly into two sub-sources. The algorithm terminates with 2 sources,
each of size 500. \rbox
\end{example}

Finally, we point out that the same techniques apply to extractors as well.
We define an extractor using the following feature vector, again ordered
from most general to most specific:
$\langle${\tt extractor, pattern, predicate, website}$\rangle$.
The finest granularity represents
the quality of a particular extractor pattern (different patterns may have
different quality), on extractions for a particular predicate (in some cases
when a pattern can extract triples of different predicates, it may have different
quality), from a particular website (a pattern may have different quality on
different websites).

\section{Experimental Results}
\label{sec:exp}
This section describes our experimental results on a synthetic data
set (where we know the ground truth), and on large-scale real-world data.
We show that (1) our algorithm can effectively estimate the correctness
of extractions, the truthfulness of triples, and the accuracy of sources; 
(2) our model significantly improves over the state-of-the-art methods 
for knowledge fusion; and (3) KBT provides a valuable additional 
signal for web source quality.

\subsection{Experiment Setup}
\subsubsection{Metrics} 
We measure how well we predict extraction correctness, triple
probability,
 and source accuracy.
For synthetic data, we have the benefit of ground truth,
so we can exactly measure all three aspects. 
We quantify this in terms of {\em square loss};
the lower the square loss, the better.
Specifically, \SqV measures the average square loss between $p(V_d=v|X)$ and the
true value of $\ind{V^*_d=v}$;
\SqC measures the average square loss between $p(C_{wdv}=1|X)$ and the true value of $\ind{C^*_{wdv}=1}$;
and \SqA measures the average square loss between $\hat{A}_w$ and the true value of $A^*_w$.

For real data, however, as we show soon, we do not have a gold standard for source
trustworthiness, and we have only a partial gold standard for triple
correctness and extraction correctness.
Hence for real data, we just focus on measuring how well we predict triple truthfulness. 
In addition to \SqV, we also used the following three metrics for this purpose,
which were also used in \cite{DGE+14}.

\begin{itemize}\tightlist
  \item {\em Weighted deviation (WDev)}: WDev measures whether the predicted probabilities are
    {\em calibrated}. We divide our triples according to the predicted probabilities
    into buckets $[0, 0.01),\\ \dots, [0.04, 0.05), [0.05, 0.1), \dots, [0.9, 0.95), 
    [0.95, 0.96), \dots,$ \\$[0.99, 1), [1, 1]$ (most triples fall in $[0,0.05)$ and $[0.95, 1]$,
    so we used a finer granularity there). For each bucket we compute the accuracy
    of the triples according to the gold standard, which can be considered as the
    real probability of the triples. WDev computes the average square loss between
    the predicted probabilities and the real probabilities, weighted by the number
    of triples in each bucket; the lower the better.
  \item {\em Area under precision recall curve (AUC-PR)}: AUC-PR measures whether the predicted probabilities are 
    {\em monotonic}. We order triples according to the computed probabilities
    and plot PR-curves, where the X-axis represents the recall and the Y-axis
    represents the precision. AUC-PR computes the area-under-the-curve;
    the higher the better. 
  \item {\em Coverage (Cov)}: Cov computes for what percentage of the triples
    we compute a probability (as we show soon, we may ignore data from a source whose quality
    remains at the default value over all the iterations).
\end{itemize}

Note that on the synthetic data Cov is 1 for all methods, and
the comparison of different methods regarding AUC-PR and WDev is very similar to
that regarding \SqV, so we skip the plots.


\subsubsection{Methods being compared}
\label{sec:methods}
We compared three main methods.
The first, which we call {\sc SingleLayer}, implements the state-of-the-art methods
for knowledge fusion~\cite{DGE+14} (overviewed in Section~\ref{sec:definition}). In particular,
each source or ``provenance'' is a 4-tuple {\tt (extractor, website, predicate, pattern)}.
We consider a provenance in fusion only if its accuracy does not remain default 
over iterations because of low coverage.
We set $n=100$ and iterate 5 times. These settings have been shown 
in~\cite{DGE+14} to perform best.

The second, which we call {\sc MultiLayer}, implements the multi-layer
model described in Section~\ref{sec:fusion}. 
To have reasonable execution time,
we used the finest granularity specified in Section~\ref{sec:hierarchy}
for extractors and sources: each extractor is an
{\tt (extractor, pattern, predicate, website)} vector, and
each source is a {\tt (website, predicate, webpage)} vector.
When we decide extraction correctness, we consider the confidence provided
by extractors, normalized to $[0,1]$,
as in Section~\ref{sec:confidenceWeightedExtractions}.
If an extractor does not provide confidence,
we assume the confidence is $1$.
When we decide triple truthfulness,
by default we use the improved estimate $p(C_{wdv}=1|X)$ 
described in Section~\ref{sec:improved},
instead of simply using $\hat{C}_{wdv}$.
We start updating the prior probabilities $p(C_{wdv}=1)$,
as described in Section~\ref{sec:reEstimatePrior},
starting from the third iteration,
since the probabilities we compute get stable after the second
iteration. 
For the noise models, we set $n=10$ and $\gamma=0.25$, 
but we found other settings lead to quite similar results.
We vary the settings and show the effect in Section~\ref{sec:exp_alt}.

The third method, which we call \SplitAndMerge, implements 
the {\sc SplitAndMerge} algorithm in addition to the multi-layer model,
as described in Section~\ref{sec:hierarchy}. 
We set the min and max sizes to $m=5$ and $M=10K$ by default,
and varied them 
in Section~\ref{sec:efficiency}.

For each method, there are two variants. The first variant determines
which version of the $p(X_{ewdv}|C_{wdv})$ model we use.
We tried both \accu and \popaccu.
We found that the performance of the two variants on the single-layer model was
very similar, while \popaccu is slightly better. However, rather surprisingly, we found that the \popaccu
version of the multi-layer model was worse than the \accu version.
This is because we have not yet found a way to combine the \popaccu
model with the improved estimation procedure described in Section~\ref{sec:improved}.
Consequently, we only report results for the \accu version in what follows.

The second variant is how we initialize source quality. We either assign a default
quality ($A_w=0.8, R_e=0.8, Q_e=0.2$) or initialize the quality according to a
gold standard, as explained in Section~\ref{sec:kvexp}. In this latter case, 
we append $+$ to the method name to distinguish it from the default initialization
(\eg, {\sc SingleLayer+}).

\subsection{Experiments on synthetic data}
\label{sec:synthetic}

\subsubsection{Data set}
\label{sec:synData}

We randomly generated data sets containing 10 sources and 5
extractors.
Each source provides 100 triples with an accuracy of $A=0.7$. 
Each extractor extracts triples from a source with probability $\delta=0.5$;
for each source, it extracts a provided triple with probability $R=0.5$;
accuracy among extracted subjects (same for predicates, objects)
is $P=0.8$ (in other words, the precision of the extractor is $P_e=P^3$). 
In each experiment we varied one parameter 
from 0.1 to 0.9 and fixed the others;
for each experiment we repeated 10 times and reported the average.
Note that our default setting represents a challenging case,
where the sources and extractors are of relatively low quality.

\subsubsection{Results}

\newcommand{\scale}{0.28}

\begin{figure*}
\vspace{-.1in}
\begin{minipage}[th]{\linewidth}
\centering
\includegraphics[scale=\scale]{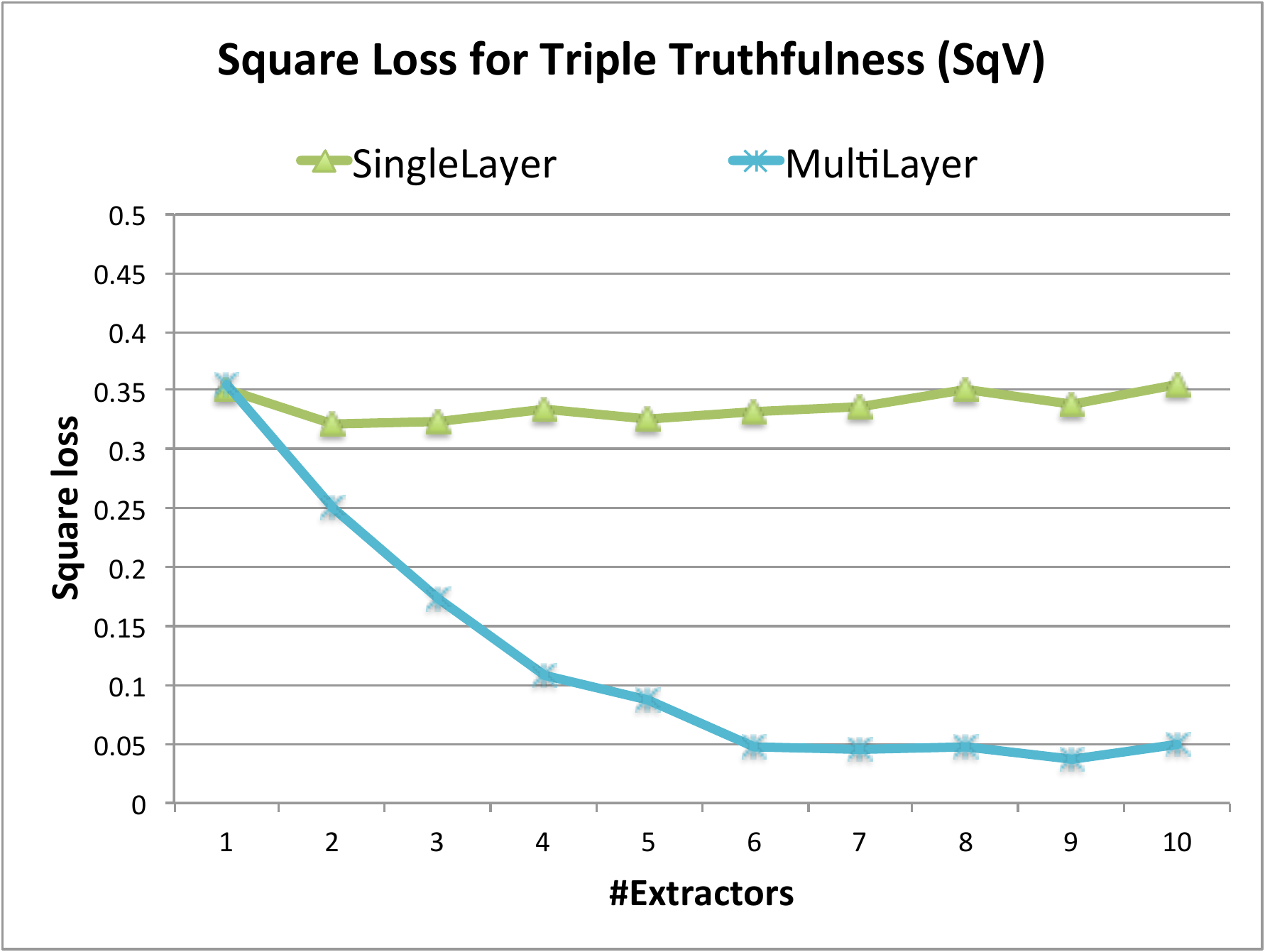}
\hfill
\includegraphics[scale=\scale]{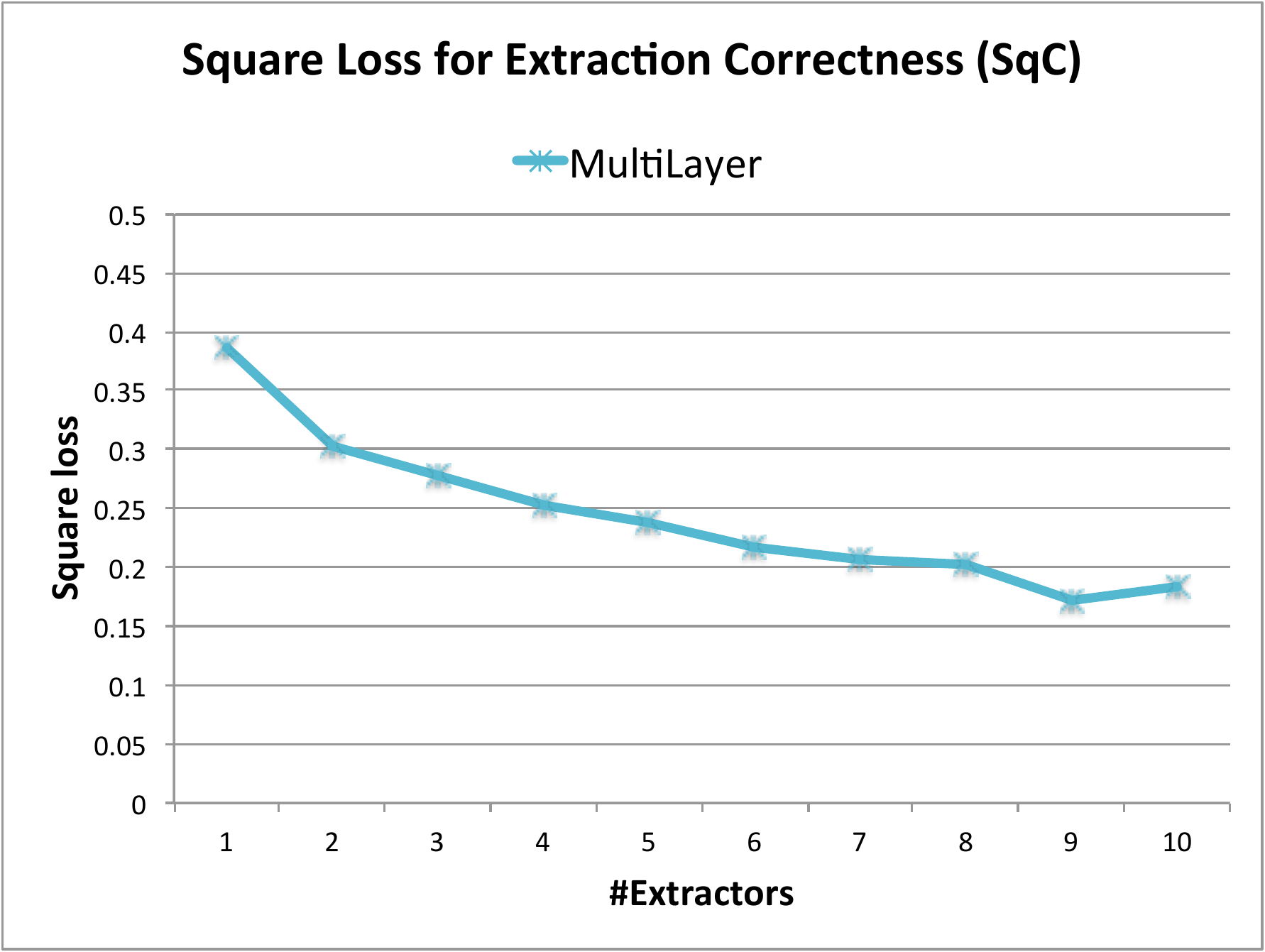}
\hfill
\includegraphics[scale=\scale]{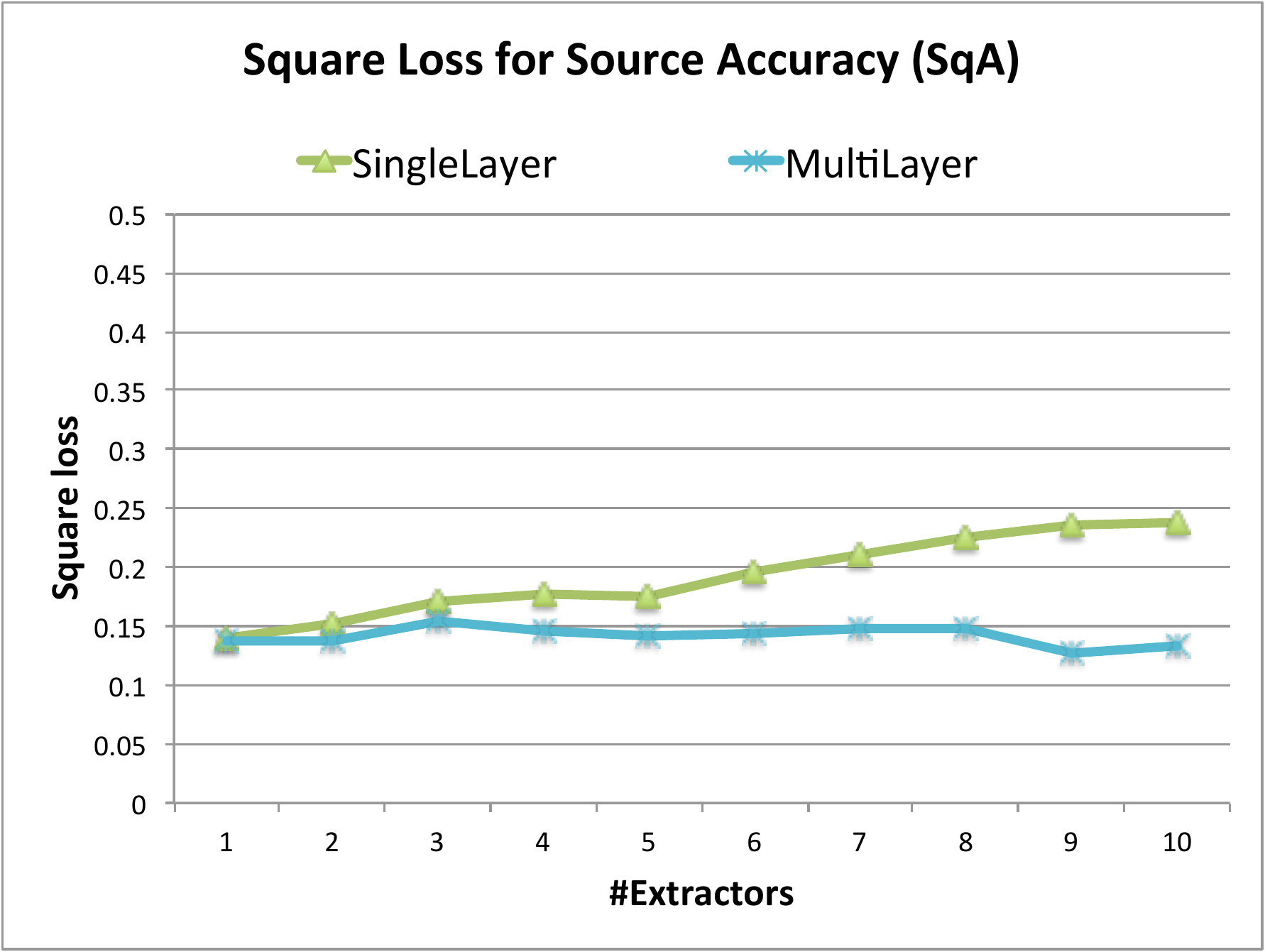}
\vspace{-.05in}
\caption{\small Error in estimating $V_d$, $C_{wdv}$ and $A_w$
as we vary the number of extractors in the synthetic data.
The multi-layer model has significantly lower square loss
than the single-layer model. The single-layer model
cannot estimate $C_{wdv}$, resulting with one line for \SqC.
\label{fig:numOfExt}}
\end{minipage}
\end{figure*}

\begin{figure*}
\begin{minipage}[th]{\linewidth}
\centering
\includegraphics[scale=\scale]{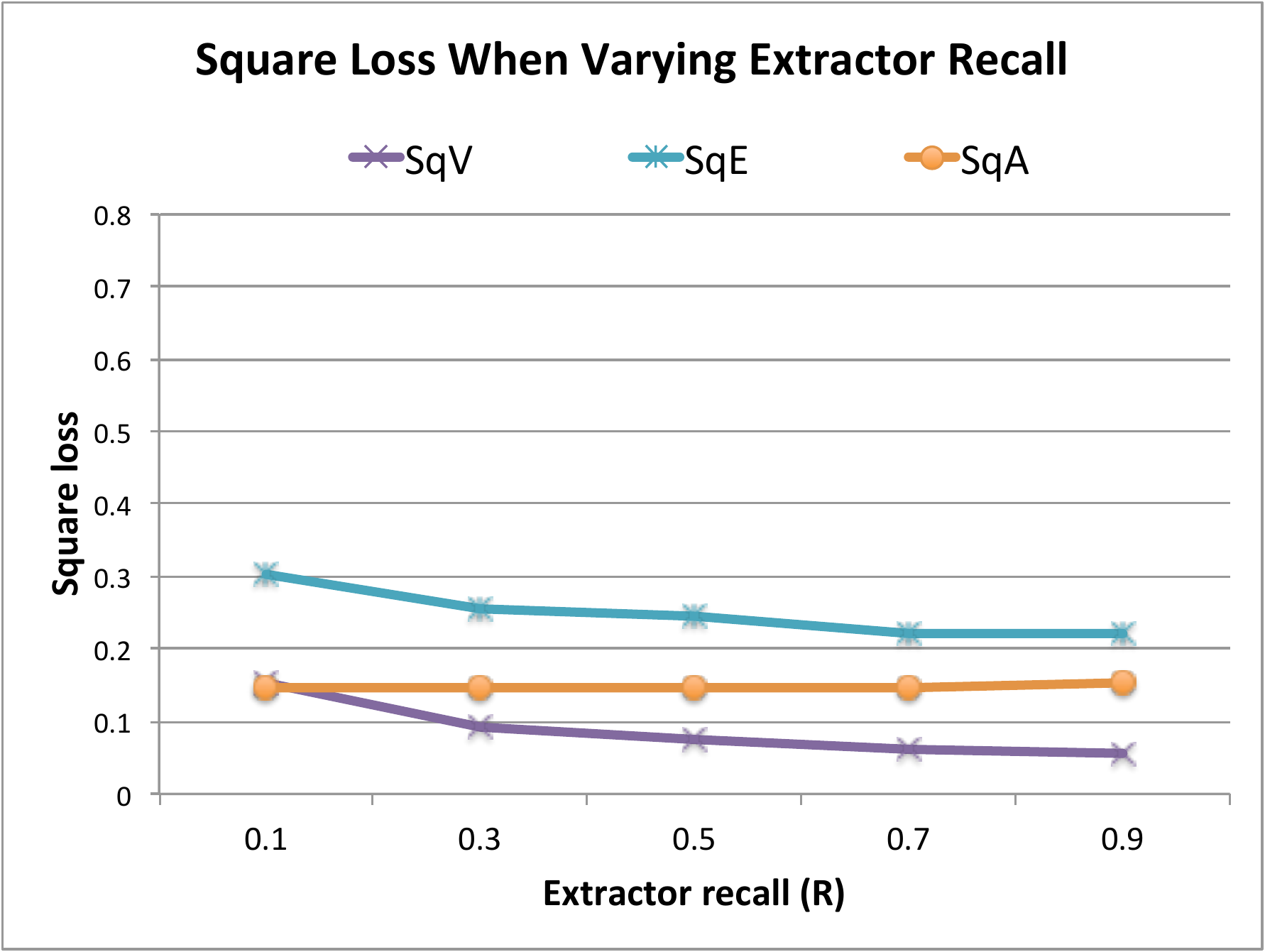}
\hfill
\includegraphics[scale=\scale]{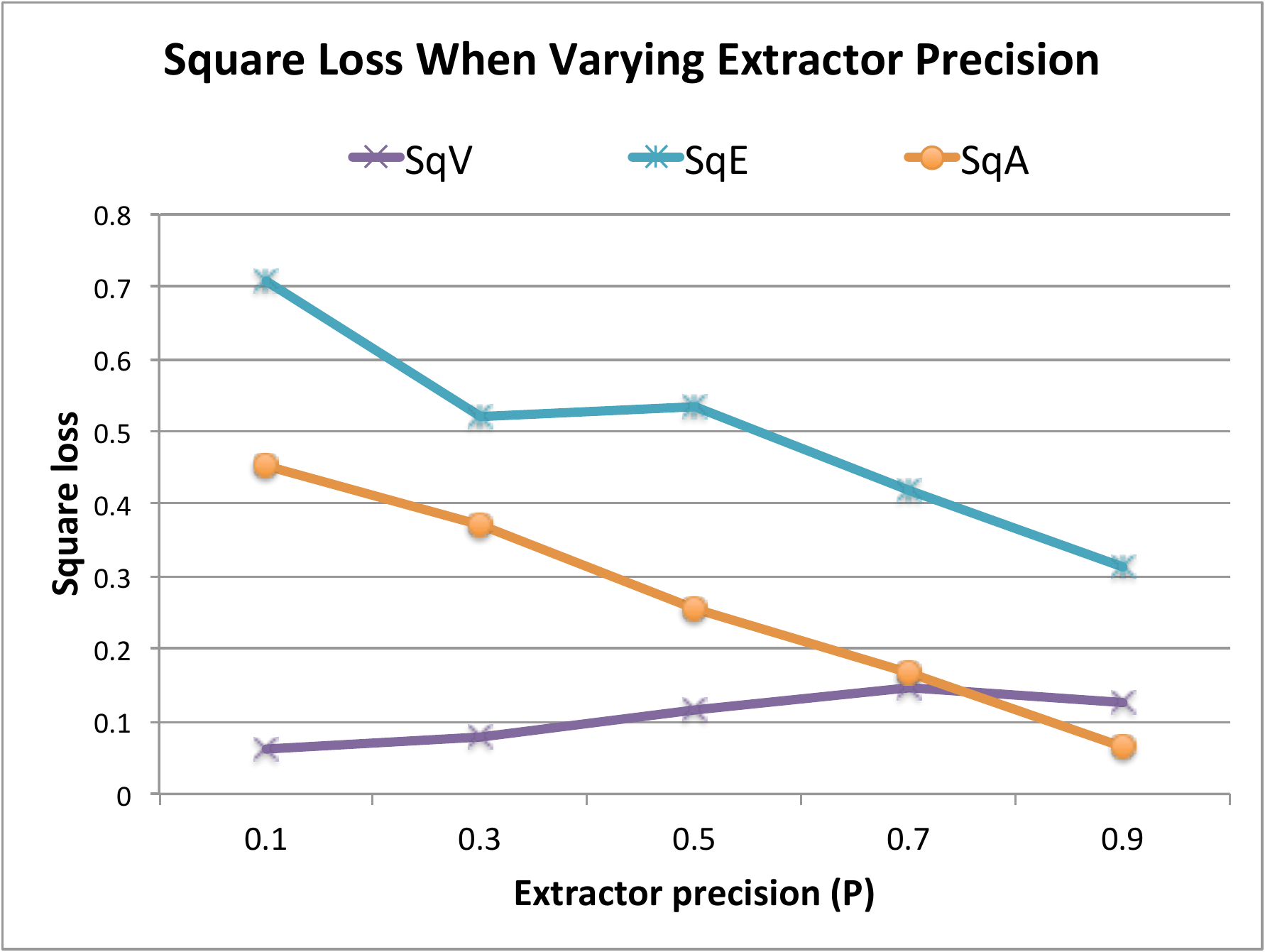}
\hfill
\includegraphics[scale=\scale]{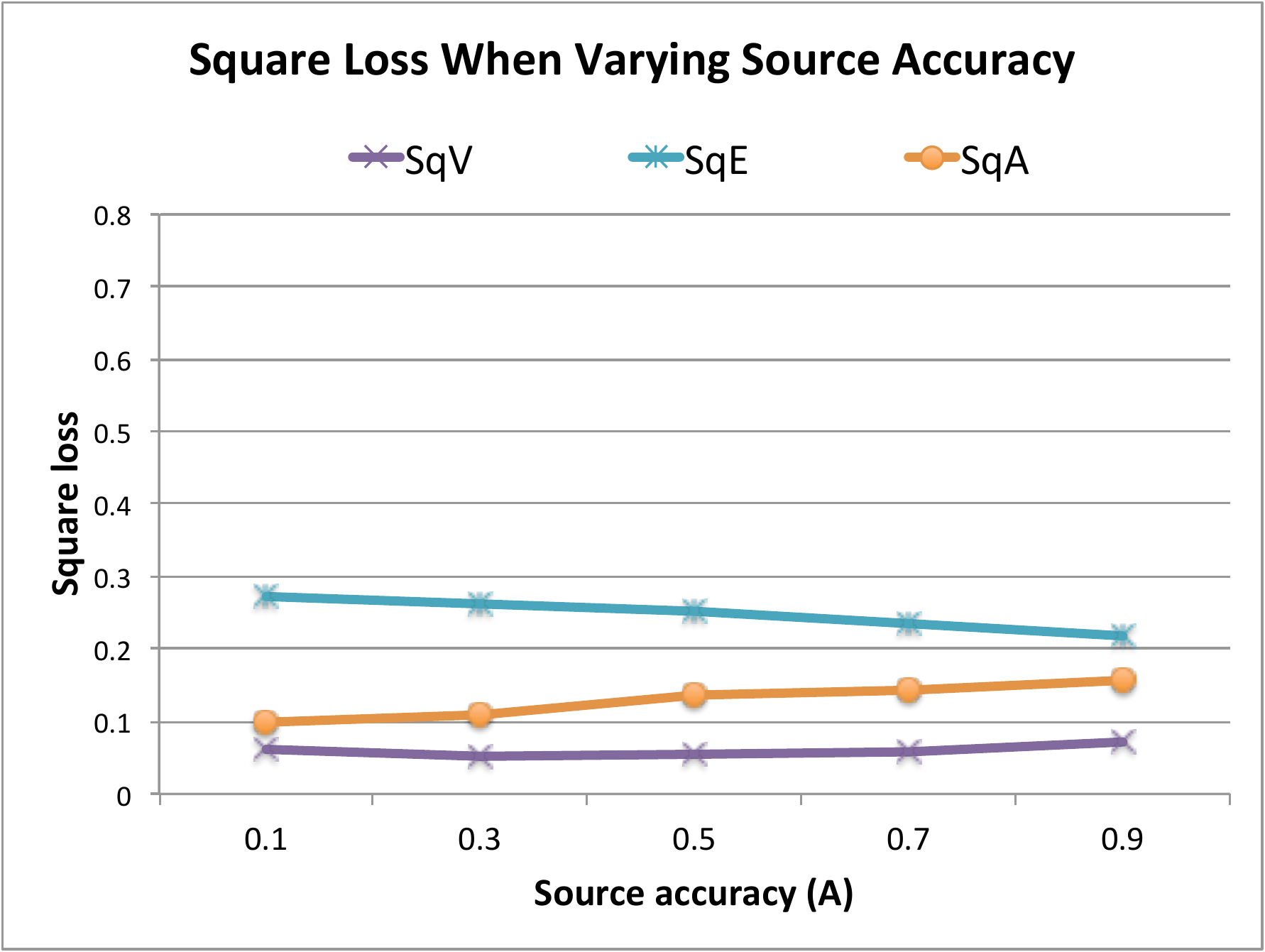}
\vspace{-.05in}
\caption{\small Error in estimating $V_d$, $C_{wdv}$ and $A_w$ as we vary extractor quality
($P$ and $R$) and source quality ($A$) in the synthetic data.
\label{fig:para}}
\end{minipage}
\vspace{-.1in}
\end{figure*}

Figure~\ref{fig:numOfExt} plots \SqV, \SqC, and \SqA
as we increase the number of extractors.
We assume {\sc SingleLayer} considers all extracted triples
when computing source accuracy.
We observe that the multi-layer model  always performs better than 
the single-layer model. 
As the number of extractors increases, \SqV goes down quickly for the
multi-layer model,
and \SqC also decreases, albeit more slowly.
Although the extra extractors can introduce much more noise extractions,
\SqA stays stable for {\sc MultiLayer}, whereas it increases quite a lot
for {\sc SingleLayer}.
\eat{ does not increase ; this is because adding more
extractors introduces more noise into the system, as well as more
parameters to estimate.}

Next we vary source and extractor quality. {\sc MultiLayer} continues
to perform better than {\sc SingleLayer} everywhere and
Figure~\ref{fig:para} plots only for {\sc MultiLayer} as we vary $R$, $P$ and $A$
(the plot for varying $\delta$ is similar to that for varying $R$).
In general the higher quality, the lower the loss.
There are a few small deviations from this trend.
When the extractor recall ($R$) increases, \SqA does not decrease, as the extractors
also introduce more noise. When the extractor precision ($P$) increases,
we give them higher trust, resulting in a slightly higher (but still low) probability for false triples;
since there are many more false triples than true ones, \SqV slightly
increases.
Similarly, when $A$ increases, there is a very slight increase in \SqA,
because we trust the false triples a bit more.
However, overall, we believe the experiments on the synthetic data
demonstrate that our algorithm is working as expected, and can
successfully approximate the true parameter values in these controlled settings.

\begin{figure*}[t]
\vspace{-.1in}
\begin{minipage}{0.32\linewidth}
\centering
\includegraphics[scale=\scale]{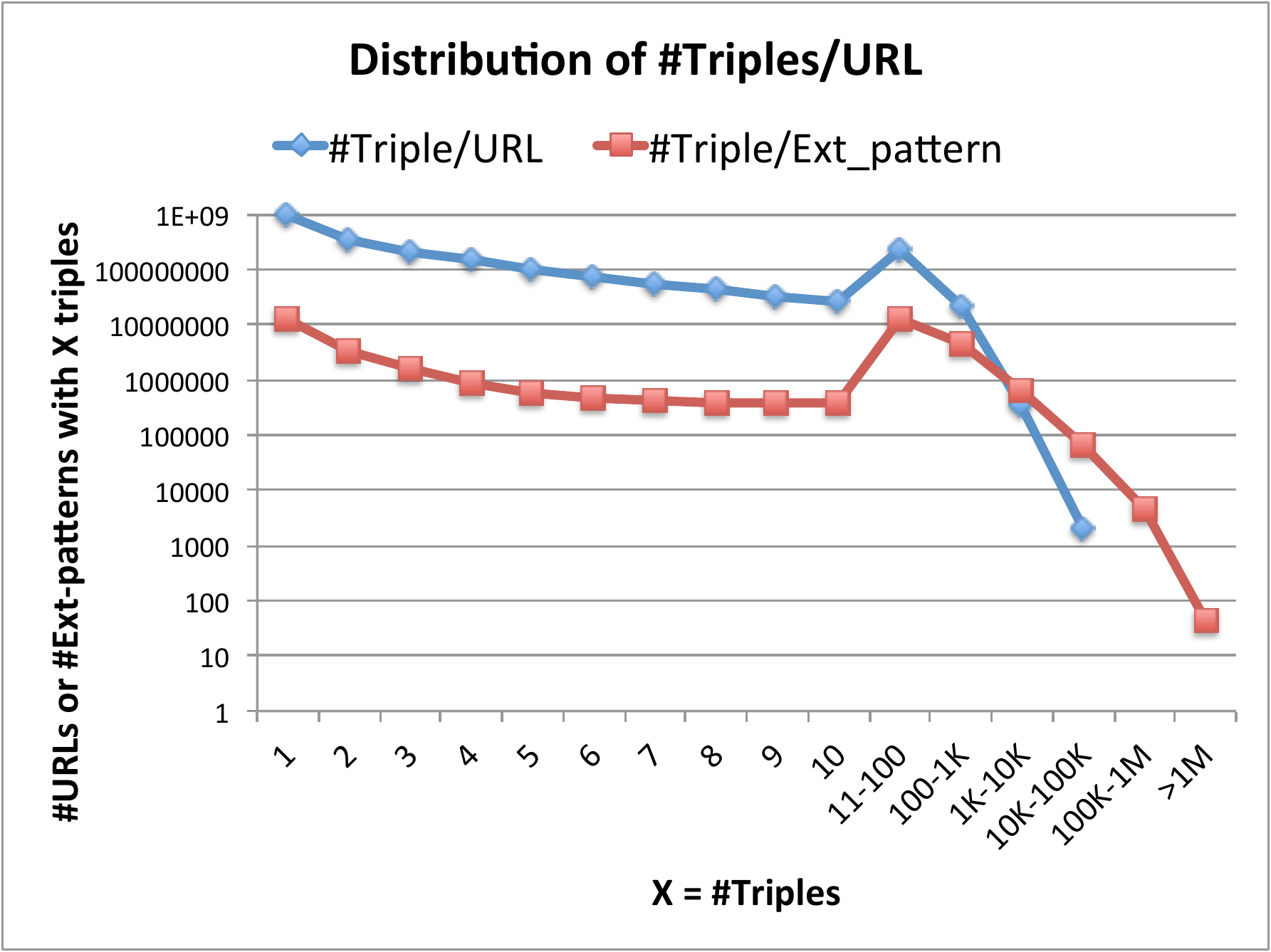}
\caption{\small Distribution of \#Triples per URL or extraction pattern
motivates {\sc SplitAndMerge}.
\label{fig:dist}}
\end{minipage}
\hfill
\begin{minipage}{0.32\linewidth}
\centering
\includegraphics[scale=\scale]{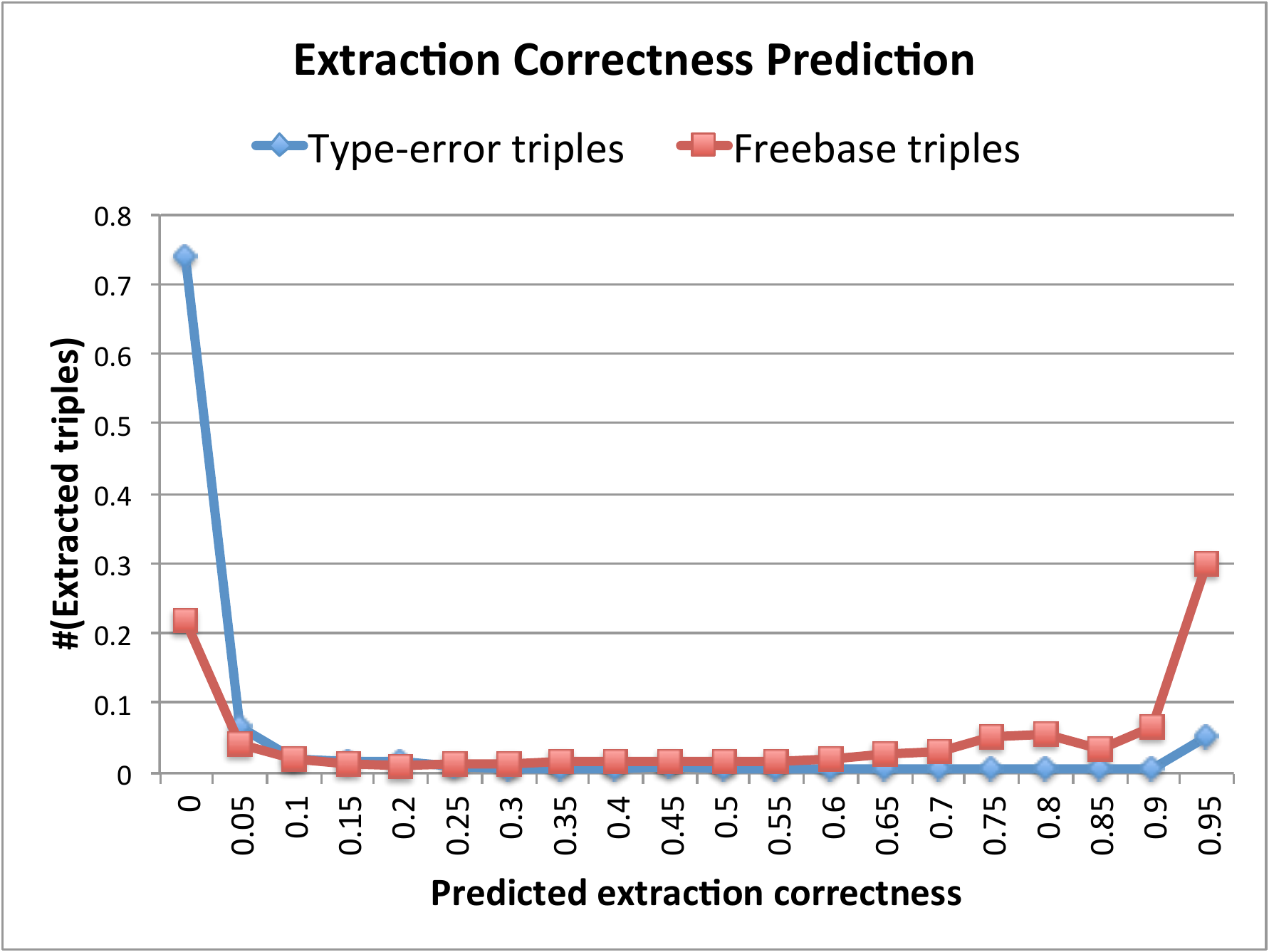}
\caption{\small Distribution of predicted extraction correctness
shows effectiveness of {\sc MultiLayer+}.
\label{fig:extCorr}}
\end{minipage}
\hfill
\begin{minipage}{0.32\linewidth}
\centering
\includegraphics[scale=\scale]{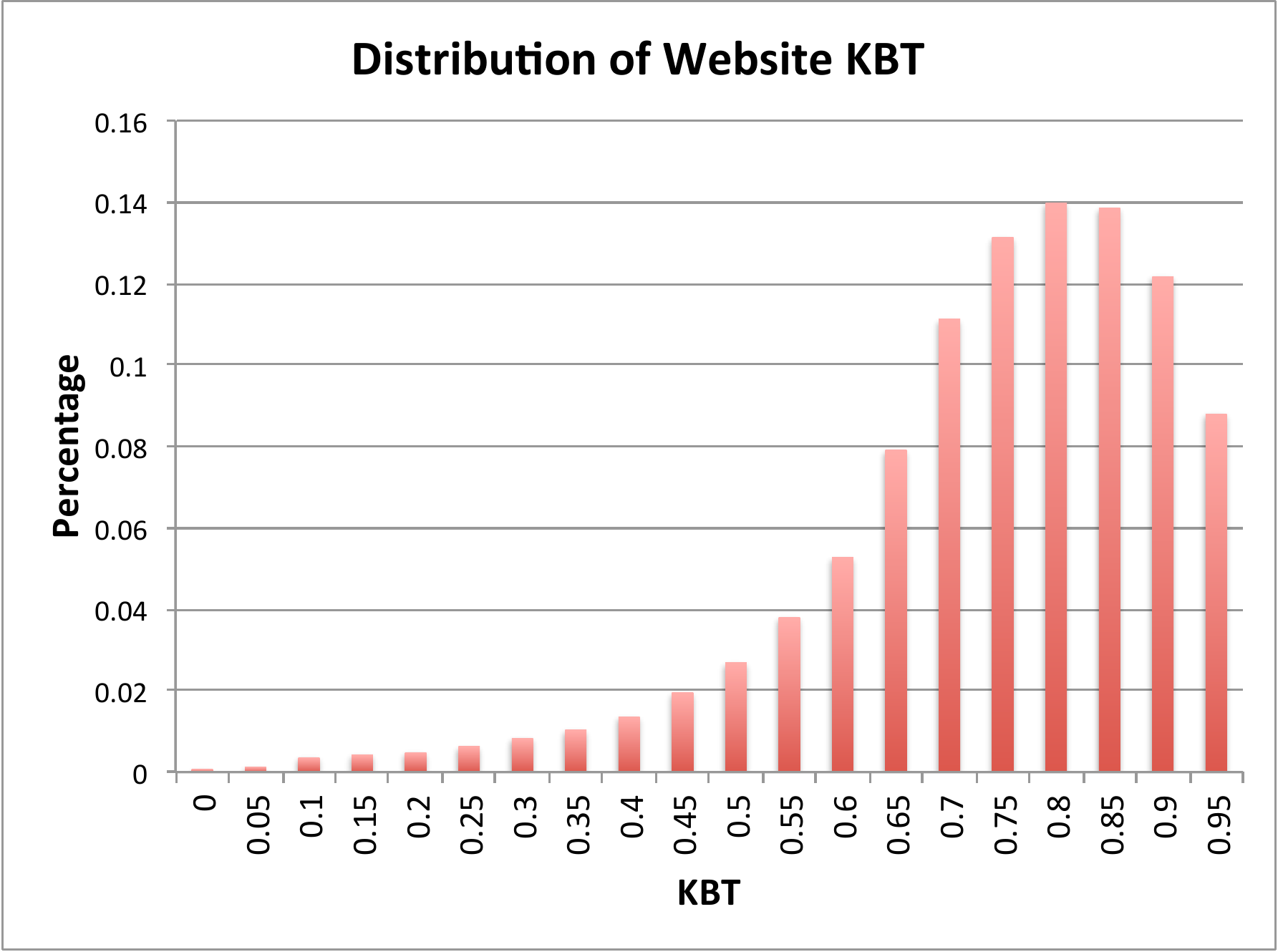}
\caption{\small Distribution on KBT for websites with at least 5
extracted triples.
\label{fig:kbt}}
\end{minipage}
\end{figure*}

\subsection{Experiments on KV data}
\label{sec:kvexp}

\subsubsection{Data set}
\label{sec:kvdata}

We experimented with knowledge triples collected by  Knowledge Vault~\cite{DGH+14} on 7/24/2014;
for simplicity we call this data set {\em KV}. 
There are 2.8B triples extracted from 2B+ webpages by 16 extractors, 
involving 40M extraction patterns. Comparing with an old version
of the data collected on 10/2/2013~\cite{DGE+14},
the current collection is 75\% larger, involves 25\% more extractors, 
8\% more extraction patterns, and twice as many webpages.

Figure~\ref{fig:dist} shows the distribution of the number of distinct extracted
triples per URL and per extraction pattern. On the one hand, we observe
some huge sources and extractors: 26 URLs each contributes over 50K triples 
(a lot due to extraction mistakes), 15 websites each contributes over
100M triples, and 43 extraction patterns each extracts over 1M triples.
On the other hand, we observe long tails: 74\% URLs each contributes fewer than 
5 triples, and 48\% extraction patterns each extracts fewer than 5 triples.
Our {\sc SplitAndMerge} strategy is exactly motivated by such observations.

To determine whether these triples are true or not (gold standard
labels),
we use two methods. The first method is called the {\em Local-Closed World 
Assumption (LCWA)}~\cite{DGH+14, DGE+14, GTH+13} and works as
follows. A triple $(s,p,o)$ 
is considered as {\tt true} if it appears in the Freebase KB.
If the triple is missing from the KB but $(s,p)$ appears for any other value $o'$,
we assume the KB is locally complete (for $(s,p)$),
and we label the $(s,p,o)$ triple as {\tt false}.
We label the rest of the triples (where $(s,p)$ is missing)
as {\tt unknown} and remove them from the evaluation set.
In this way we can decide truthfulness 
of 0.74B triples (26\% in {\em KV}), of which 20\% are true
(in Freebase).

Second, we apply type checking to find incorrect extractions.
In particular, we consider a triple $(s,p,o)$ as {\tt false} if 
1) $s=o$; 2) the type of $s$ or $o$ is incompatible with what is required
by the predicate; or 3) $o$ is outside the expected range (\eg, the weight
of an athlete is over 1000 pounds). We discovered 0.56B triples (20\% in KV) that violate
such rules and consider them both as {\tt false} triples and as extraction mistakes.

Our gold standard include triples from both labeling methods. It contains in total 1.3B
triples, among which 11.5\% are true. 

\begin{figure*}[t]
\begin{minipage}{0.32\linewidth}
\centering
\includegraphics[scale=\scale]{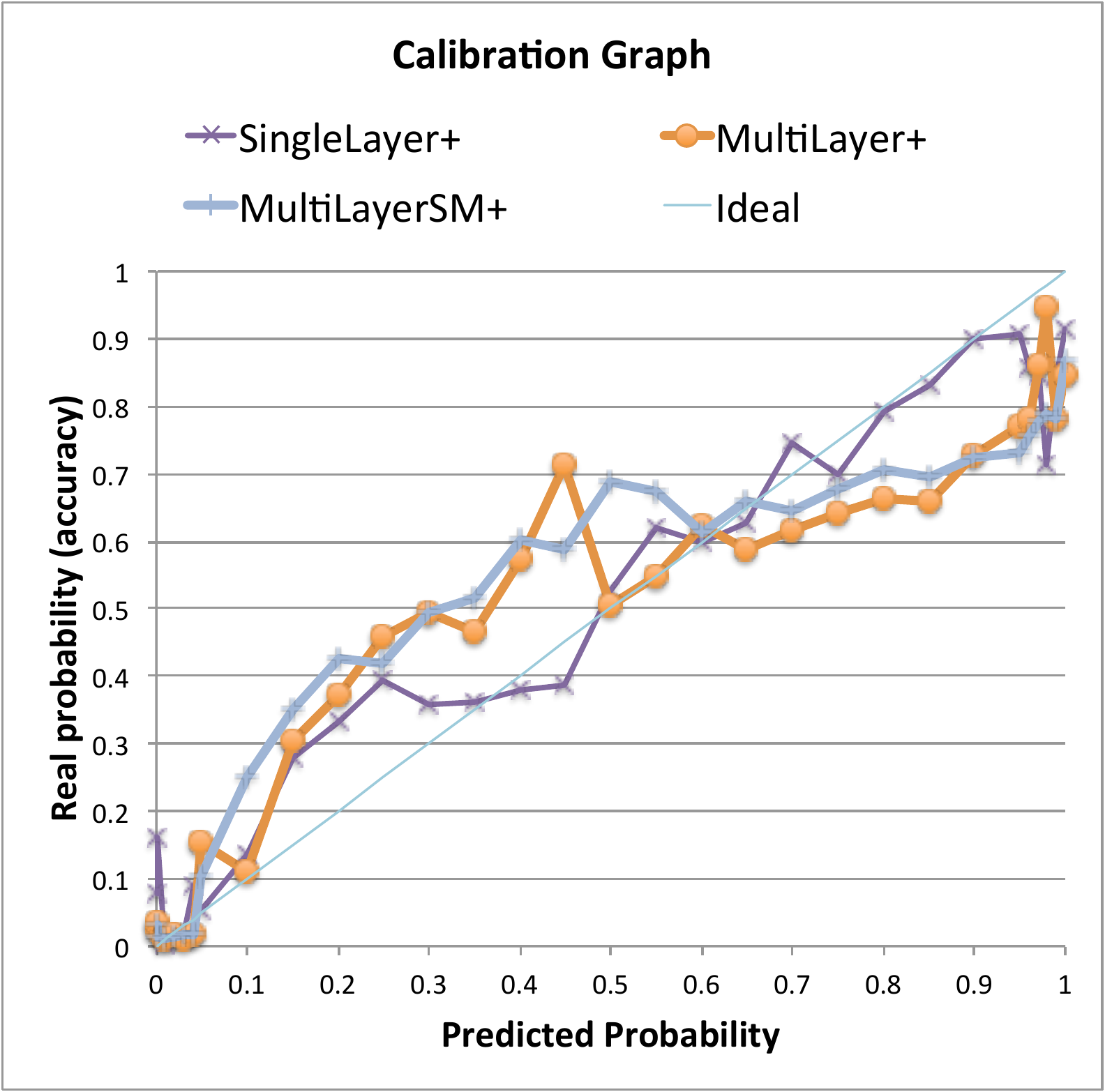}
\caption{\small Calibration curves for various methods on KV data.
\label{fig:cali}}
\end{minipage}
\hfill
\begin{minipage}{0.32\linewidth}
\centering
\includegraphics[scale=\scale]{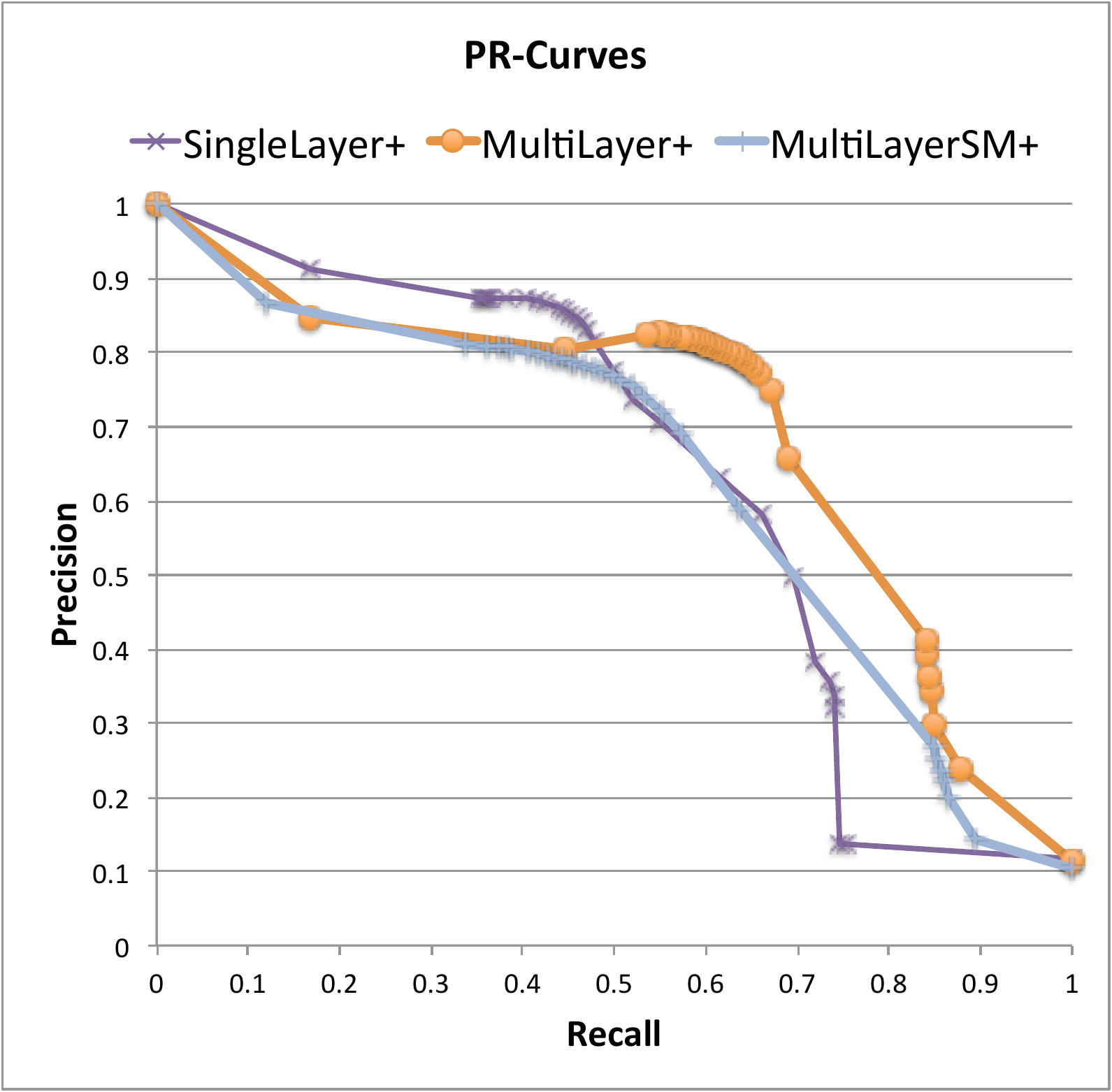}
\caption{\small PR-curves for various methods on KV data. 
{\sc MultiLayer+} has the best curve.
\label{fig:pr}}
\end{minipage}
\hfill
\begin{minipage}{0.32\linewidth}
\centering
\vspace{-.05in}
\includegraphics[scale=\scale]{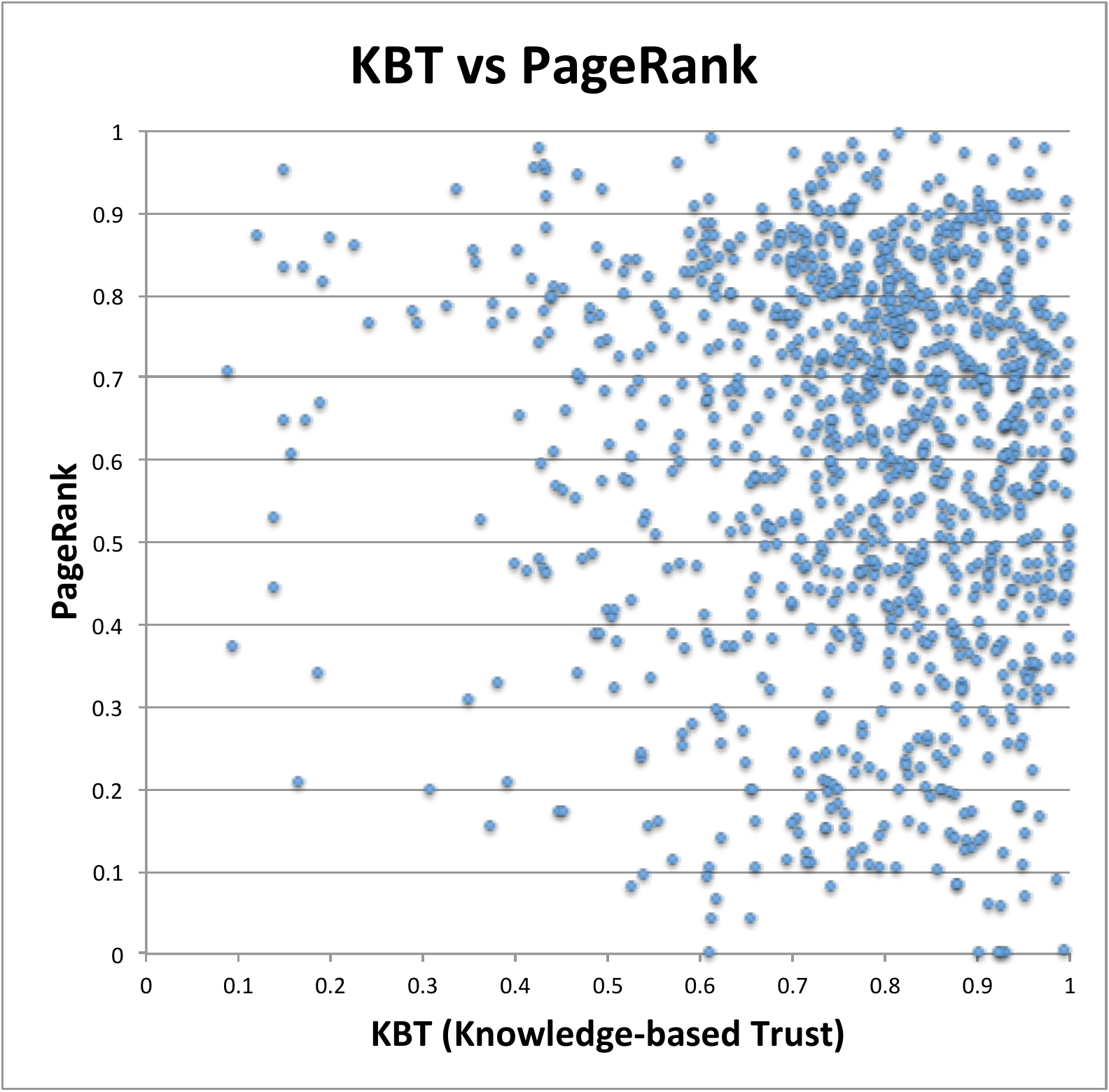}
\caption{\small KBT and PageRank are orthogonal signals.
\label{fig:correlation}}
\end{minipage}
\vspace{-.15in}
\end{figure*}

\subsubsection{Single-layer vs multi-layer}
\label{sec:overall}

{\small
\begin{table}[t]
\center
\caption{Comparison of various methods on {\em KV};
best performance in each group is in bold.
For \SqV and WDev, lower is better;
for AUC-PR and Cov, higher is better.
\label{tbl:overall}}
\vspace{-.1in}
\begin{tabular}{|c||c|c||c|c|}
\hline
\SqV & \SqV & WDev & AUC-PR & Cov \\
\hline
{\sc SingleLayer} & 0.131 & 0.061 & {\bf 0.454} & {\bf 0.952} \\
{\sc MultiLayer} & 0.105 & 0.042 & 0.439 & 0.849 \\
\SplitAndMerge & {\bf 0.090} & {\bf 0.021} & 0.449 & 0.939 \\
\hline
{\sc SingleLayer+} & 0.063 & 0.0043 & 0.630 & 0.953 \\
{\sc MultiLayer+} & {\bf 0.054} & 0.0040 & {\bf 0.693} & 0.864 \\
\SplitAndMergePlus & 0.059 & {\bf 0.0039} & 0.631 & {\bf 0.955} \\
\hline
\end{tabular}
\vspace{-.2in}
\end{table}
}

Table~\ref{tbl:overall} compares the performance of the three methods.
Figure~\ref{fig:cali} plots the calibration curve and Figure~\ref{fig:pr}
plots the PR-curve.
We see that all methods are fairly well calibrated, but the multi-layer 
model has a better PR curve. In particular, {\sc SingleLayer} often
predicts a low probability for true triples and hence has a lot of false negatives.

We see that 
\SplitAndMerge has better results than {\sc MultiLayer},
but surprisingly,
\SplitAndMergePlus has lower performance than {\sc MultiLayer+}.
That is, there is an interaction between the granularity of the
sources and the way we initialize their accuracy.

The reason for this is as follows.
When we initialize source and extractor quality using default values,
we are using unsupervised learning (no labeled data).
In this regime, \SplitAndMerge merges small sources 
so it can better predict their quality,
which is why it is better than standard {\sc MultiLayer}.
Now consider when 
we initialize source and
extractor quality using the gold standard; in this case, we are essentially using semi-supervised learning.
Smart initialization helps the most when we use a fine
granularity for sources and extractors, since in such cases we often have
much fewer data for a source or an extractor.

Finally, to examine the quality of our prediction on extraction correctness
(recall that we lack a full gold standard), we plotted
the distribution of the predictions on triples
with type errors (ideally we wish to predict a probability of 0 for them)
and on correct triples (presumably a lot of them, though not all, would be correctly
extracted and we shall predict a high probability).
Figure~\ref{fig:extCorr} shows the results by {\sc MultiLayer+}. 
We observe that for the triples with type errors, 
 {\sc MultiLayer+} predicts a probability below 0.1
for 80\% of them and a probability above 0.7 for only 8\%;
in contrast, for the correct triples in Freebase,
{\sc MultiLayer+} predicts a probability below 0.1
for 26\% of them and a probability above 0.7 for 54\%,
showing effectiveness of our model.

\subsubsection{Effects of varying the inference algorithm}
\label{sec:exp_alt}

{\small
\begin{table}[t]
\center
\caption{Contribution of different components, where significantly
  worse values (compared to the baseline)
are shown in italics.
\label{tbl:components}}
\vspace{-.1in}
\begin{tabular}{|c||c|c||c|c|}
\hline
\SqV & \SqV & WDev & AUC-PR & Cov \\
\hline
{\sc MultiLayer+} & 0.054 & 0.0040 & 0.693 & 0.864 \\
\hline
$p(V_d|\hat{C}_{d})$ & {\em 0.061} & 0.0038 & {\em 0.570} & 0.880 \\
Not updating $\alpha$ & 0.055 & {\em 0.0057} & 0.699 & 0.864 \\
$p(C_{dwv}|\ind{\overline{X}_{ewdv} > \phi})$ & 0.053 & 0.0040 & 0.696 & 0.864\\
\hline
\end{tabular}
\vspace{-.1in}
\end{table}
}

Table~\ref{tbl:components} shows the effect of changing different
pieces of the multi-layer inference algorithm, as follows.

Row $p(V_d|\hat{C}_{d})$ shows the change we incur by treating $C_d$ as
  observed data when inferring $V_d$ (as described in Section~\ref{sec:estV}),
as opposed to using the confidence-weighted version
in Section~\ref{sec:improved}.
We see a significant drop in the AUC-PR metric and an
increase in \SqV by ignoring uncertainty in $C_d$;
indeed, we predict a probability below 0.05 for the truthfulness of 93\% triples.

Row ``Not updating $\alpha$'' shows the change we incur
if we keep $p(C_{wdv}=1)$ fixed at $\alpha$,
as opposed to using the updating scheme described in 
Section~\ref{sec:reEstimatePrior}.
We see that most metrics are the same, but WDev has gotten significantly worse,
showing that the probabilities are less well calibrated.
It turns out that not updating the prior often results in
over-confidence when computing $p(V_d|X)$,
as shown in Example~\ref{ex:prior}.

Row $p(C_{dwv}|\ind{\overline{X}_{ewdv} > \phi})$  shows the change we
     incur by thresholding the confidence-weighted extractions at a
     threshold of $\phi=0$, as opposed to using the
     confidence-weighted extension in
     Section~\ref{sec:confidenceWeightedExtractions}.
 Rather surprisingly, we see that thresholding seems to work slightly better;
 however, this is consistent with previous observations that some extractors
 can be bad at predicting confidence~\cite{DGE+14}.

\subsubsection{Computational efficiency}
\label{sec:efficiency}

{\small
\begin{table}[t]
\center
\caption{
Relative running time, where 
we consider one iteration of {\sc MultiLayer}
as taking $1$ unit of time.
We see that using split and split-merge is, on average,
3 times faster per iteration.
\label{tbl:efficiency}}
\vspace{-.1in}
\begin{tabular}{|c|c|c|c|c|}
\hline
\multicolumn{2}{|c|}{Task} & Normal & Split & Split\&Merge \\
\hline
            & Source & 0 & 0.28 & 0.5 \\
Prep. & Extractor & 0 & 0.50 & 0.46 \\
\cline{2-5}
            & {\em Total} & {\em 0} & {\em 0.779} & {\em 1.034} \\
\hline
            & I. ExtCorr & 0.097 & 0.098 & 0.094 \\
            & II. TriplePr & 0.098 & 0.079 & 0.087 \\
Iter.  & III. SrcAccu & 0.105 & 0.080 & 0.074 \\
            & IV. ExtQuality & 0.700 & 0.082 & 0.074 \\
\cline{2-5}
            & {\em Total} & {\em 1} & {\em 0.337} & {\em 0.329} \\
\hline
\multicolumn{2}{|c|}{Total} & 5 & 2.466 & 2.679 \\
\hline
\end{tabular}
\vspace{-.2in}
\end{table}
}

All the algorithms were implemented in FlumeJava~\cite{CRP+10}, which
is based on Map-Reduce.
Absolute running times can vary dramatically depending on how many
machines we use. Therefore,
Table~\ref{tbl:efficiency} shows only the relative efficiency of the algorithms. 
We reported the time for preparation, including applying splitting
and merging on web sources and on extractors; and the time for
iteration, including computing extraction correctness,
computing triple truthfulness, computing source accuracy,
and computing extractor quality. For each component in the iterations,
we report the average execution time among the five iterations. 
By default $m=5, M=10K$.

First, we observe that splitting large sources and extractors can significantly
reduce execution time. In our data set some extractors extract a huge number
of triples from some websites. Splitting such extractors has a speedup
of 8.8 for extractor-quality computation. In addition, we observe that 
splitting large sources also reduces execution time by 20\% for source-accuracy
computation. On average each iteration has a speed up of 3. Although there is
some overhead for splitting, the overall execution time dropped by half.

Second, we observe that applying merging in addition does not add much overhead.
Although it increases preparation
by 33\%, it drops the execution time in each iteration slightly (by 2.4\%) because
there are fewer sources and extractors. The overall execution
time increases over splitting by only 8.6\%. Instead, a baseline strategy that
starts with the coarsest granularity and then splits big sources and extractors
slows down preparation by 3.8 times.

Finally, we examined the effect of the $m$ and $M$ parameters. We observe that varying $M$ from 1K to 50K 
affects prediction quality very little; however, setting $M=1K$ (more splitting) slows down preparation
by 19\% and setting $M=50K$ (less splitting) slows down the inference by 21\%, so both have longer
execution time. On the other hand, increasing $m$ to be above 5 does not change the performance much,
while setting $m=2$ (less merging) increases wDev by 29\% and slows down inference by 14\%.

\eat{
\begin{figure}[t]
\center
\vspace{-.1in}
\includegraphics[scale=0.5]{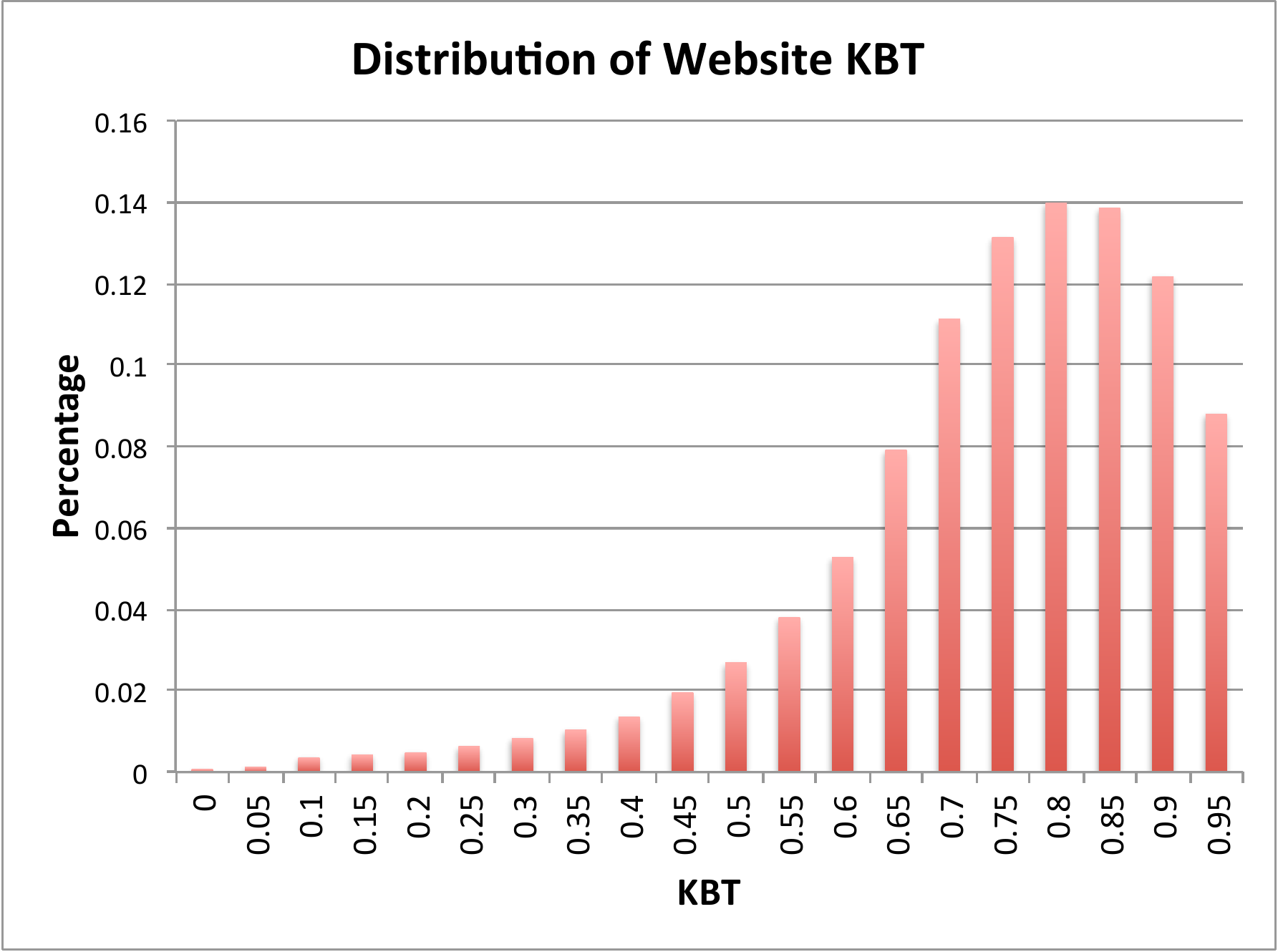}
\caption{Distribution on KBT for websites.
\label{fig:kbt}}
\vspace{-.1in}
\end{figure}
}
\subsection{Experiments related to KBT}
\label{sec:expKBT}

We now evaluate how well we estimate the trustworthiness of webpages.
Our data set contains 2B+ webpages from 26M websites. 
Among them, our multi-layer model believes that we have correctly
extracted at least 5 triples from about 119M webpages and 5.6M websites. 
Figure~\ref{fig:kbt} shows the distribution of KBT scores:
we observed that the peak is at 0.8 and 52\% of the websites have a KBT over 0.8.

\subsubsection{KBT vs PageRank}
Since we do not have ground truth on webpage quality,
we compare our method to PageRank.
We compute PageRank for all webpages on the web, and normalize the scores to $[0,1]$.
Figure~\ref{fig:correlation} plots KBT and PageRank for 2000 randomly selected websites. 
As expected, the two signals are almost orthogonal.
We next investigate the two cases where KBT differs significantly from PageRank.

\smallskip
\noindent
{\bf Low PageRank but high KBT (bottom-right corner):} To understand which sources may obtain high KBT,
we randomly sampled 100 websites whose KBT is above 0.9. The number of extracted
triples from each website varies from hundreds to millions. 
For each website we considered the top 3 predicates and randomly selected 
from these predicates 10 triples where the probability of the extraction 
being correct is above 0.8.
We manually evaluated each website according to the following 4 criteria.

\vspace{-.05in}
\begin{itemize}\tightlist
  \item {\em Triple correctness}: whether at least 9 triples
    are correct.
  \item {\em Extraction correctness}: whether at least 9 triples
    are correctly extracted (and hence we can evaluate the website according to
    what it really states).
  \item {\em Topic relevance}: we decide the major topics 
    for the website according to the website name and the introduction
    in the ``About us'' page; we then decide whether at least 9 triples are relevant
    to these topics (\eg, if the website is about business directories in
    South America but the extractions are about cities and countries
    in SA, we consider them as not topic relevant).
  \item {\em Non-trivialness}: we decide whether the sampled triples
    state non-trivial facts (\eg, if most sampled triples from a Hindi movie website
    state that the language of the movie is Hindi, we consider it as trivial).
\end{itemize}
\vspace{-.1in}

We consider a website as truly trustworthy if it satisfies all 
of the four criteria. Among the 100 websites, 85 are considered trustworthy;
2 are not topic relevant, 12 do not have enough non-trivial triples,
and 2 have more than 1 extraction errors (one website has two issues).
However, only 20 out of the 85 trustworthy sites have a PageRank over 0.5. 
This shows that KBT 
can identify sources with trustworthy data, even though they are tail
sources with low PageRanks.

\smallskip
\noindent
{\bf High PageRank but low KBT (top-left corner):} 
We consider the 15 gossip websites listed in~\cite{gossip}.
Among them, 14 have a PageRank among top 15\% of the websites,
since such websites are often popular.
However, for all of them the KBT are in the bottom 50\%;
in other words, they are considered less trustworthy than half
of the websites.
Another kind of websites that often get low KBT are forum websites.
For instance, we discovered that {\em answers.yahoo.com} 
says that {\em ``Catherine Zeta-Jones is from New Zealand''}
\footnote{\small https://answers.yahoo.com/question/index?qid=20070206090808AAC54nH.},
although she was born in Wales according to 
{\em Wikipedia}\footnote{\small http://en.wikipedia.org/wiki/Catherine\_Zeta-Jones.}.

\subsubsection{Discussion}
\label{sec:disc}

Although we have seen that KBT seems to provide a useful signal about
trustworthiness, which  is orthogonal to more traditional signals such
as PageRank, our experiments also show places for further improvement as 
future work.

\vspace{-.05in}
\begin{enumerate}\tightlist
  \item To avoid evaluating KBT on topic irrelevant triples,
    we need to identify the main topics
    of a website, and filter triples whose entity or 
    predicate is not relevant to these topics.

  \item To avoid evaluating KBT on trivial extracted triples,
    we need to decide whether the information in a triple is trivial.
    One possibility is to consider a predicate
    with a very low variety of objects as less informative.
    Another possibility is to associate triples with an 
IDF (inverse document frequency), such that low-IDF triples
get less weight in KBT computation.

  \item Our extractors (and most state-of-the-art extractors)
    still have limited extraction capabilities and this limits
    our ability to estimate
    KBT for all websites. We wish to increase our KBT coverage by
    extending our method to handle open-IE style information
    extraction techniques, which do not conform to a schema~\cite{Etzioni11}.
However, although these methods
can extract more triples, they may introduce more noise.

  \item Some websites scrape data from other websites. Identifying such websites
    requires techniques such as copy detection. Scaling up copy
    detection techniques, such as \cite{DBH+10a, DBS09a}, has been attempted in \cite{LDL+15},
but more work is required before
these methods can be applied to analyzing extracted data from billions of web sources.

\eat{
  \item Finally, there have been many other signals such as PageRank, visit history,
    spaminess for evaluating web source quality. 
    Combining KBT with those signals would be important future work.
}
\end{enumerate}
\vspace{-.05in}

\section{Related Work}
\label{sec:related}
There has been a lot of work studying how to assess quality of web sources.
PageRank~\cite{pagerank} and Authority-hub analysis~\cite{Kleinberg98}
consider signals from link analysis (surveyed in~\cite{BRRT05}). 
EigenTrust~\cite{KSG03} and TrustMe~\cite{SL03} consider signals from
source behavior in a P2P network. Web topology~\cite{CDG+07},
TrustRank~\cite{GGP04}, and AntiTrust~\cite{KR06} detect web spams.
The knowledge-based trustworthiness we propose in this paper is different 
from all of them in that it considers an important {\em endogenous}
signal --- the 
correctness of the factual information provided by a web source.

Our work is relevant to the body of work in {\em Data fusion}
(surveyed in~\cite{BN08, DN09, LDL+15}),
where the goal is to resolve conflicts from data provided by multiple sources
and find the truths that are consistent with the real world.
Most of the recent work in this area considers trustworthiness of sources,
measured by link-based measures~\cite{PR10, PR11}, IR-based measures~\cite{WM07},
accuracy-based measures~\cite{DBS09a, DBS09b, DSS13, LLG+14, PDD+14, YHY07},
and graphical-model analysis~\cite{PR13, YT11, ZRHG12, ZH12}.
However, these papers do not model 
the concept of an extractor, and hence they cannot distinguish an
unreliable source from an unreliable extractor.

\eat{
Our multi-layer fusion model extends from two basic models in data fusion:
one for extraction correctness~\cite{PDD+14} and one for triple 
truthfulness~\cite{DBS09a}. However, our major contribution in this paper is 
the multi-layer fusion framework that enables evaluation of website 
trustworthiness. In addition, we extended the two models in three ways
to make them fit in multi-layer fusion. First, we incorporate 
the probabilities that may be associated with the input data.
Second, we consider a better way of setting {\em a priori} probabilities. 
Third, we discuss how we may adjust the granularity level of sources dynamically according
to the sizes and the hierarchy of the sources. The first two extensions 
allow seamless propagation of the uncertainty from one layer to the other layer,
and the last extension allows better quality estimation.
Our experimental results have shown how each of the extensions helps in
improving the results (Section~\ref{sec:kvexp}).
}

Graphical models have been proposed to solve the data fusion 
problem~\cite{PR13, YT11, ZH12, ZRHG12}. These models are more or less
similar to our single-layer model in Section~\ref{sec:singleLayer};
in particular, \cite{PR13} considers single truth, \cite{ZH12} considers
numerical values, \cite{ZRHG12} allows multiple truths, and \cite{YT11}
considers correlations between the sources. However, these prior works do not model 
the concept of an extractor,  
and hence they cannot capture the fact 
that sources and extractors introduce qualitatively different kinds of noise. 
In addition, the data sets used in their experiments are typically 5-6
orders of magnitude smaller in scale than ours, 
and their inference algorithms are inherently slower than our algorithm.
The multi-layer model and the scale of our experimental data also 
distinguish our work from other data fusion techniques.


Finally, the most relevant work is our previous work on knowledge fusion~\cite{DGE+14}.
We have given detailed comparison in Section~\ref{sec:overview},
as well as empirical comparison in Section~\ref{sec:exp},
showing that {\sc MultiLayer} improves over {\sc SingleLayer} for knowledge
fusion and gives the opportunity of evaluating KBT for web source quality.
\eat{
in our previous work~\cite{DGE+14} we proposed the 
knowledge fusion problem as an important step in knowledge curation;
extensive data analysis and experimental study have been conducted 
to evaluate the effectiveness of adapting data-fusion models 
as a single-layer model in solving this more challenging problem~\cite{DGE+14}. 
This paper proposes a multi-layer model that distinguishes 
triple correctness and extraction correctness, and separates
source accuracy and extractor quality (precision and recall).
We have compared our techniques with the solutions in~\cite{DGE+14}
in detail in Section~\ref{sec:overview} and empirically shown in Section~\ref{sec:exp}
that {\sc MultiLayer} improves over {\sc SingleLayer} for knowledge
fusion and gives the opportunity of evaluating KBT for web-source quality.
}

\section{Conclusions}
\label{sec:conclude}
This paper proposes a new metric for evaluating web-source 
quality--knowledge-based trust.
We proposed a sophisticated probabilistic model that jointly
estimates the correctness of extractions and source data,
and the trustworthiness of sources. In addition, we presented
an algorithm that dynamically decides the level
of granularity for each source. Experimental results have shown 
both promise in evaluating web source quality and improvement 
over existing techniques for knowledge fusion.

\eat{
\section{Acknowledgements}
We thank Fernando Pereira, Divesh Srivastava, and Amar Subramanya for
constructive suggestions that helped us improve this paper.
We also thank Anish Das Sarma, Alon Halevy, Kevin Lerman, Abhijit Mahabal,
and Oksana Yakhnenko for help with the extraction pipeline.
}

\balance

{\small
\bibliographystyle{abbrv}
\bibliography{base}
}

\end{document}